\UseRawInputEncoding
\documentclass[reprint,superscriptaddress,amsmath,amssymb,amsfonts,aps,prb,floatfix,longbibliography]{revtex4-2}
\usepackage{times}
\usepackage{graphicx}
\usepackage{dcolumn}
\usepackage{bm}
\usepackage{ulem}
\usepackage[colorlinks=true, citecolor=blue, linkcolor=blue, urlcolor=blue]{hyperref}
\usepackage{bbold}
\DeclareMathAlphabet\mathbfcal{OMS}{cmsy}{b}{n}
\usepackage{wasysym}
\usepackage{amssymb}
\usepackage{amsmath}
\usepackage[usenames,dvipsnames]{xcolor}

\begin{document}

\title{Inducing Superconductivity in Bilayer Graphene by Alleviation of the Stoner Blockade}
\author{Gal Shavit}
\affiliation{Department of Condensed Matter Physics,    Weizmann Institute of Science, Rehovot, Israel 7610001}

\author{Yuval Oreg}
\affiliation{Department of Condensed Matter Physics,    Weizmann Institute of Science, Rehovot, Israel 7610001}

\date{\today}

\begin{abstract}
External magnetic fields conventionally suppress superconductivity, both by orbital and paramagnetic effects.
A recent experiment has shown that in a Bernal stacked bilayer graphene system, the opposite occurs -- a finite critical magnetic field is necessary to observe superconducting features occurring in the vicinity of a magnetic phase transition.
We propose an extraordinary electronic-correlation-driven mechanism by which this anomalous superconductivity manifests.
Specifically, the electrons tend to avoid band occupations near high density of states regions due to their mutual repulsion.
Considering the nature of spontaneous symmetry breaking involved, we dub this avoidance \textit{Stoner blockade}.
We show how a magnetic field softens this blockade, allowing weak superconductivity to take place, consistent with experimental findings.
Our principle prediction is that a small reduction of the Coulomb repulsion would result in sizable superconductivity gains, both in achieving higher critical temperatures and expanding the superconducting regime.
Within the theory we present, magnetic field and spin-orbit coupling of the Ising type have a similar effect on the  Bernal stacked bilayer graphene system, elucidating the emergence of superconductivity when the system is proximitized to a $\rm WSe_2$ substrate.
We further demonstrate in this paper the sensitivity of superconductivity to disorder in the proposed scenario. We find that a disorder that does not violate Anderson's theorem may still induce a reduction of $T_c$ through its effect on the density of states, establishing the delicate nature of the Bernal bilayer graphene superconductor.

\end{abstract}

\maketitle

A superconductor subject to an external magnetic field usually suffers deterioration of its superconducting properties: the superconducting gap and transition temperature are suppressed, vortices are introduced into the bulk of the material, and resistivity increases~\cite{GinzburgLandau,tinkham2004introduction}.
The magnetic field's most harmful aspect is its orbital effect on the superconducting condensate.
This effect can be almost entirely suppressed when the magnetic field is applied parallel to thin films (whose width is much smaller than the London penetration depth) or in two-dimensional materials, e.g., graphene.

Yet, in these materials, the magnetic field's adverse effect may persist in the form of pair breaking due to the Zeeman effect.
Namely, if the electron pairs that make up the superconducting condensate have opposite spins, then the Zeeman coupling to their spin magnetic moment eventually eliminates superconductivity.
In the case of conventional spin-singlet superconductors, the Pauli-Chandrasekhar-Clogston limit~\cite{Clogston,ClogsonLimit} sets the critical field strength at $\Delta/\left(\sqrt{2}\mu_B\right)$ ($\Delta$ is the superconducting gap, $\mu_B$ is the Bohr magneton).

In recent years superconductors that are not very sensitive to magnetic fields have emerged.
These materials are very thin, up to a single atomic layer, and have a non-singlet superconducting order parameter facilitated by their multi-orbital band structure and electronic correlations.
The most notable examples are few-layer transition metal dichalcogenides, presumably hosting so-called Ising superconductivity~\cite{IsingSuperconductivityMoS2,IsingSuperconductivityNbSe2,delaBarrera2018IsingSuperconductivityTas2Nbse2TUNING}, and twisted graphene multilayers~\cite{Cao2021PauliLimitViolationTrilayer,Park2022MATngFamily,NadjPergAscendenceMATng}.

A recent experiment~\cite{ZhouYoungBLGZeeman} has discovered an even more extreme example of the effect of magnetic fields.
Remarkably, the authors found that in electrically-biased Bernal-stacked bilayer graphene (BLG), superconductivity emerges in the hole-doped side of the charge neutrality point only \textit{above} a critical in-plane magnetic field strength (which also exceeds the Pauli limiting field).
This material's superconducting regime appears to lie close to a magnetic phase transition, making the phenomena even more peculiar.

We present and study the following scenario as a possible explanation of the magnetic-field-induced superconductivity in BLG. In the absence of an external magnetic field, an electrical displacement field modifies the BLG (non-interacting) band structure such that the density may be tuned to the vicinity of a van-Hove singularity (vHS) with a large density of states (DOS). However, when Coulomb interactions between the electrons are introduced, due to the large DOS, a Stoner-like phase transition occurs so that some bands are occupied more than others. In this spontaneously reconstructed distribution of the occupations, non of the bands is near the vHS, and the interaction energy is minimized. 

We find that applying an external parallel magnetic field weakens this ``Stoner Blockade'' effect, allowing the system to park near configurations with a larger DOS. 
Analyzing a first-order phase transition under general considerations, we find that this is a generic outcome to be expected  when applying a field that couples to the order parameter.
The presence of the large normal-state DOS enables in turn, the stabilization of a superconducting phase, whose $T_c$ is large enough to be observed experimentally.  

Thus, our theory gives rise to superconductivity residing exactly around the phase transition line, as is experimentally observed.
A straightforward prediction of the theory we present is that a slight suppression of the Coulomb repulsion by, e.g., tuning the strength of screening by a nearby metallic gate (cf. Refs.~\cite{YoungTuningSC,EfetovTuningSC,BLGscreening}), can lead to a dramatic expansion of the parameter regime supporting superconductivity.

The novel Stoner blockade mechanism we present has two additional appealing features hinting at its relevance to BLG.
First, it easily generalizes to the scenario where the in-plane field is replaced by an Ising spin orbit coupling (ISOC) term in the band structure.
It thus accounts for some of the phenomenology found in other experiments~\cite{ZhangBLGSOC,BLGyoungNadjperge}, where enhanced superconductivity was measured in BLG proximate to a $\rm WSe_2$ monolayer.
Second, we demonstrate that within this framework, due to the required high DOS in our scenario, only pristine high mobility stat-of-the-art devices would display superconducting behavior, even in the presence of protection by the so-called Anderson's theorem~\cite{ANDERSONtheorem}.
This somewhat resolves the issue of the scarcity of superconducting BLG devices to date, requiring recent major advances in device quality.

The rest of the manuscript is organized as follows.
In Sec.~\ref{sec:stonerblockade} we describe how electron interactions give rise to a forbidden range of Fermi-level energies close to the vHS within a simple Hartree-Fock picture. 
We sketch how this can be detrimental to superconductivity and how an in-plane magnetic field partially alleviates the blockade.
The superconductivity calculations, taking into account the instantaneous Coulomb repulsion and a retarded pairing mechanism, are described in Sec.~\ref{sec:superconductivitycalculations}. 
The residual pair-breaking orbital effect of the magnetic field is also considered.
The case of ISOC and the importance of (non-pair-breaking) disorder are discussed in Sec.~\ref{sec:refinements}.
Finally, we conclude our discussion in Sec.~\ref{sec:conclusions}, and comment on several open questions.

\section{Stoner blockade}\label{sec:stonerblockade}

\subsection{Normal state ``cascade''}\label{sec:normalstatecascade}
In this work we focus on studying the Hamiltonian
\begin{equation}
    H = H_0 +H_{\rm int} +{\cal H}_{\rm SB},\label{eq:normalstateHamiltoniangeneral}
\end{equation}
where $H_0$ describes the low-energy dispersion of electrons in BLG, $H_{\rm int}$ is a phenomenological short-range interaction Hamiltonian, and ${\cal H}_{\rm SB}$ is an $SU\left(2\right)$ symmetry-breaking operator to be discussed later on.
We define $\Psi_{\mathbf k}$ as an 8-spinor of fermionic annihilation operators at momentum $\mathbf k$, with pseudo-spin (layer), valley, and spin degrees of freedom, described by Pauli matrices $\sigma_i$, $\tau_i$, and $s_i$, respectively.

The single particle Hamiltonian may be written as~\cite{MccanBernalGrapheneLandauLevel,McCannBilayerGrapheneBernal},
\begin{equation}
    H_0 = \sum_{\mathbf k} \Psi_{\mathbf k}^\dagger\left(
    h_0 + h_{\rm tri} + h_{\rm Dis} + h_{\rm p.h.}
    \right)\Psi_{\mathbf k}.\label{eq:singleparticleh0}
\end{equation}
Here, the matrix $h_0$ accounts for the quadratic band touching in each valley, $h_{\rm tri}$ describes the trigonal warping due to sub-leading interlayer tunneling, $h_{\rm Dis}$ describes the potential difference between the layers induced by an electric displacement field, and $h_{\rm p.h.}$ accounts for particle-hole asymmetric terms.
The different terms are given by
\begin{subequations}\label{eq:detailsofsinglebandterms}
    \begin{equation}
        h_0=-\frac{v^2}{\gamma_1}\left[\left(k_x^2-k_y^2\right)\sigma_x+2k_xk_y\sigma_y\tau_z\right],\label{eq:quadbandtouching}
    \end{equation}
    \begin{equation}
        h_{\rm tri} = v_3\left(k_x \sigma_x\tau_z - k_y\sigma_y\right),\label{eq:trigonalwarping}
    \end{equation}
    \begin{equation}
        h_{\rm Dis} = - U{\left( 1 - 2 \frac{v^2 k^2}{\gamma_1^2}\right)}\sigma_z, \label{eq:displacementfield}
    \end{equation}
    \begin{equation}
        h_{\rm p.h.} = \left(2\frac{v v_4}{\gamma_1} +\Delta' \frac{v^2}{\gamma_1^2}\right)k^2.\label{eq:particleholeasymmetry}
    \end{equation}
\end{subequations}
In the expressions above $k^2 = k_x^2 +k_y^2$, and $2U$ is the potential difference between the graphene layers. 
Here we use the parameters $v=1.1\times10^6\frac{\rm m}{\rm sec}$, $v_3=1.3\times10^5\frac{\rm m}{\rm sec}$, $v_4=4.8\times10^4\frac{\rm m}{\rm sec}$, $\gamma_1=381\,{\rm meV}$,  and $\Delta'=22\,{\rm meV}$.

\begin{figure}
\begin{centering}
\includegraphics[scale=0.6,viewport=0bp -5bp 450bp 702bp]{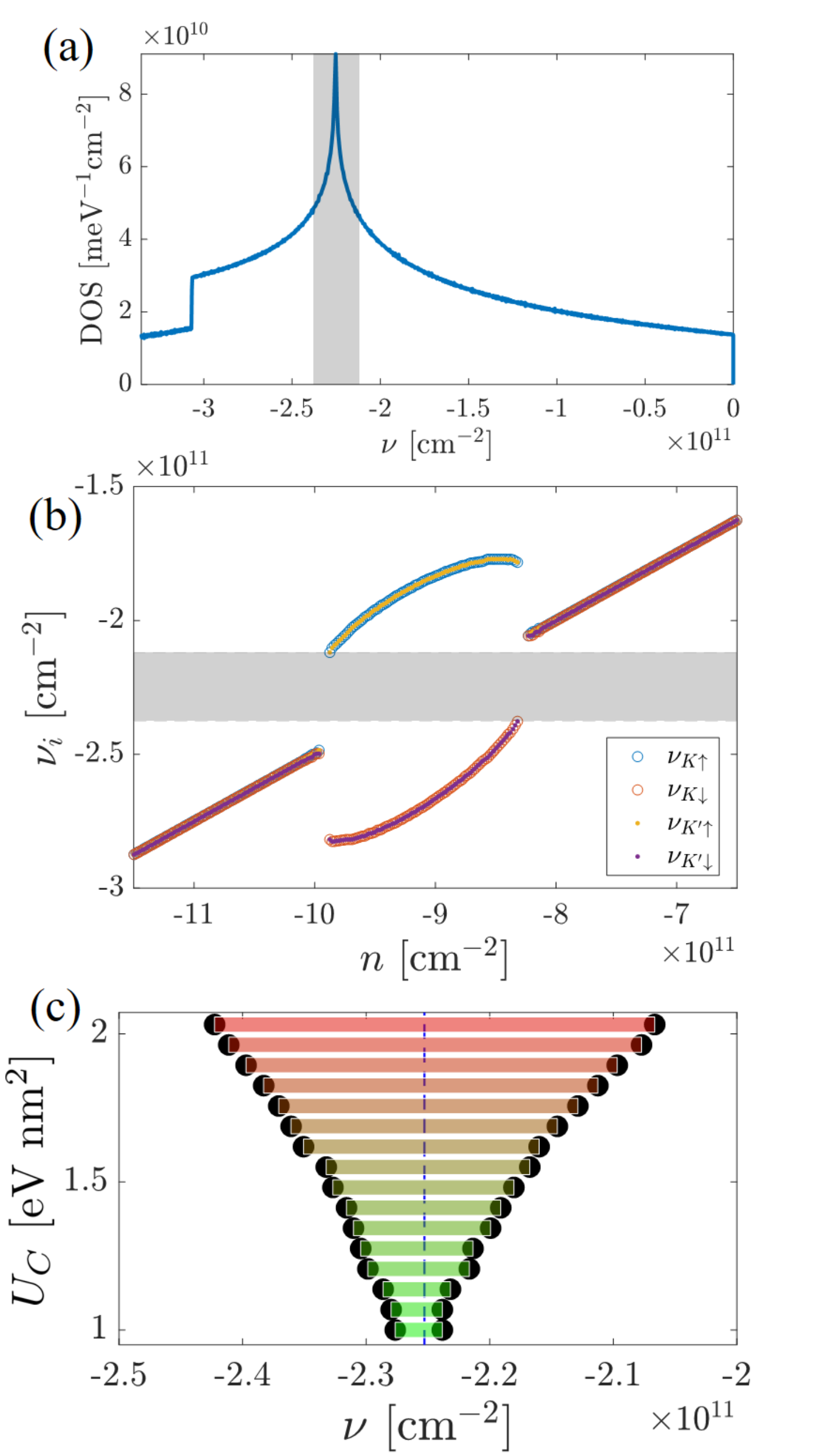}
\par\end{centering}
\caption{ \label{fig:StonerBlockade}
(a) 
Density of states per flavor of the valence band, computed from the Hamiltonian $H_0$ [Eq.~\eqref{eq:singleparticleh0}]. 
The grey rectangle demarcates the blockaded region in panel (b). 
(b)
Flavor resolved densities $\nu_i$ as a function of total electron filling, calculated by the variational Hartree-Fock method.
Spontaneous spin polarization develops in the system approaching the van-Hove filling from either side.
The gray rectangle emphasizes the forbidden range of flavor density due to the strong electronic interactions.
Here, $U_C=1.8$~eV$\,$nm$^2$, $J=0.25$~eV$\,$nm$^2$.
(c)
Extent of the Stoner blockade with varying interaction strength $U_C$.
The van Hove filling is marked by a dashed blue line.
Throughout this figure we use $U=60$~meV.
}
\end{figure}

In the presence of a large displacement field a gap opens in the band structure at charge neutrality, and the DOS features pronounced van-Hove singularities.
An example of the valence band DOS, (which will be our focus since it is where superconductivity was observed in experiments) is shown in Fig.~\ref{fig:StonerBlockade}a.

Next, electronic interactions in our Hamiltonian are given by
\begin{equation}
    H_{\rm int}\!=\!\frac{1}{\Omega}
    \sum_{\mathbf q}\left(
    \frac{U_C}{2}N_{\mathbf q}N_{-{\mathbf q}} 
    +U_{V} n^+_{\mathbf{q}}n^-_{\mathbf{-q}}
    + J\mathbf{S}^+_{{\mathbf q}}\cdot\mathbf{S}^-_{-{\mathbf q}} 
    \right)    \label{eq:Hintterm},
\end{equation}
where
$N_{\mathbf q}=\sum_{\mathbf k}\Psi^\dagger_ {\mathbf {k+q}}\Psi_ {\mathbf {k}}$,
$n^{\pm}_{\mathbf q}=\sum_{\mathbf k}\Psi^\dagger_ {\mathbf {k+q}}\frac{1\pm\tau_z}{2}\Psi_ {\mathbf {k}}$,
$\mathbf{S}^\pm_{{\mathbf q}}=\sum_{\mathbf k}\Psi^\dagger_ {\mathbf {k+q}}\frac{1\pm\tau_z}{2}\mathbf{s}\Psi_ {\mathbf {k}}$,
and $\Omega$ is the system area.
The structure of $H_{\rm int}$ is the most general form of short-range interactions which respect the symmetry of the system: time-reversal,  $SU\!\left(2\right)$ spin symmetry (in the absence of magnetic fields or spin-orbit coupling), and the $U\!\left(1\right)$ charge and (approximate) valley symmetries.
The interaction term proportional to $U_C$ is a structure-less density-density interaction, which is entirely $SU\!\left(4\right)$ symmetric in valley-spin space, and is considered to be dominant as compared to the other two terms.
The term proportional to $U_V$ accounts for possible differences between intravalley and intervalley density-density interactions and will be set to zero throughout this work, as it is non-essential for correctly capturing the phenomenology we aim to study.
Finally, $J$ is the intervalley Hund's coupling between electron spins in opposite valleys.
The experimental phenomenology in hBN-encapsulated BLG~\cite{ZhouYoungBLGZeeman} (also in rhombohedral trilayer graphene~\cite{Zhou2021RTGYoung}) is most consistent with the Hund's interaction being ferromagnetic, i.e., $J<0$~\footnote{The shifting phase boundary as a function of magnetic field and density seen in Ref.~\cite{ZhouYoungBLGZeeman} is suggestive of linear coupling between the magnetic field and the order parameter (see Sec.~\ref{sec:softeningtransitionsmagnetization} and Fig.~\ref{fig:softenedtransition}a).
As such, the possibility of spontaneous spin-polarization and Zeeman coupling of the electrons to the in-plane magnetic field is naturally the most appealing.
 }.

We may now analyze the model of Eq.~\eqref{eq:normalstateHamiltoniangeneral} using a variational Hartree-Fock approach, similar to the the ones employed in Refs.~\cite{DiracRevivals,ShavitMaATBGprl,Zhou2021RTGYoung}.
Our interest lies on the hole-doped side of charge neutrality in the system, where the peculiar superconducting phenomenon was experimentally observed.
Moreover, for the physical effect illustrated in this work it is sufficient to consider flavor symmetry broken phases, i.e., order parameters which are some combination of $\tau_z$ and $s_z$ alone.

Our analysis thus proceeds as follows.
At a given chemical potential $\mu$, the grand-potential 
$\Phi=\left\langle H - \mu N_0\right\rangle_{\rm H.F.}$ is minimized,
where $\left\langle\right\rangle_{\rm H.F.}$ denotes the expectation value calculated using the variational wavefunction
\begin{equation}
    |\Psi\rangle_{\rm H.F.}= \prod_i \left(\prod_{\substack{{\mathbf k} \\ \epsilon_{\mathbf k}>\mu_i}} c_{i,{\mathbf k}}\right)|{\rm CN}\rangle, \label{eq:variationalwavefunctionpsi}
\end{equation}
where $c_{i,{\mathbf k}}$ annihilates an electron of flavor $i$ in the valence band at momentum $\mathbf k$ with energy $\epsilon_{\mathbf k}$, $|{\rm CN}\rangle$ is the flavor-symmetric charge-neutral Fermi-sea, and $\mu_i\leq 0$ are the four variational parameters corresponding to the four spin-valley flavors, with the index $i=\left(\tau,s\right)$ combining the spin and valley indices (see Appendix~\ref{app:HartreeFock}).
Obtaining the different $\mu_i$, we calculate the flavor resolved densities 
$\nu_i=-\frac{1}{\Omega}
\sum_{\substack{{\mathbf k} \\ \mu_i<\epsilon_{\mathbf k}<0}}$,
and their relation to the total density $n=\sum_i \nu_i$.

We focus on the vicinity of the valence band vHS in the high displacement field regime, where the anomalous superconductivity was observed.
Fig.~\ref{fig:StonerBlockade}b demonstrated the typical pattern of phase transitions we observe with a reasonable choice of parameters, consistent with the experimental picture. 
The system favors a flavor-symmetric phase at lower hole densities, transitions into a two-fold symmetric spontaneously spin-polarized phase with increased doping, and then becomes flavor symmetric again as more holes are doped into it.

The electrons spontaneously develops flavor polarization as to avoid an energetically unfavorable phase, where the Fermi levels of all the bands sit at a high DOS region.
The interaction energy cost of such a phase, given that $U_C$ is strong enough, triggers a Stoner-like transition.
The ferromagnetic intervalley Hund's interaction $J$ is responsible for the specific pattern of flavor polarization, where valley degeneracy is preserved, yet the electrons spin polarize.
We note that the phase transitions are first-order, i.e., the magnetization is discontinuous across the transition.
This is consistent with the weakly negative compressibility measured at the vicinity of this transition~\cite{YoungPerrgeBLGcompressibility}, generally associated with first-order transitions accompanied by phase separation.

Tracking the evolution of the individual flavor densities, one notices an interesting feature.
The flavors tend to avoid certain densities, which encompass the vHS (Fig.~\ref{fig:StonerBlockade}a).
We term this interaction-induced blocking of certain flavor-resolved densities the \textit{Stoner blockade}.
Unsurprisingly, the extent of the blockaded region is directly related to the strength of repulsive interactions, as demonstrated in Fig.~\ref{fig:StonerBlockade}c.

\subsection{Blocking superconductivity}\label{sec:blockingsuperconductivity}
Let us briefly discuss the implications of the demonstrated Stoner blockade (detailed calculations of superconductivity within our model are carried out in Sec.~\ref{sec:superconductivitycalculations}).
The experimentally measured critical temperature of the superconducting phase, of order  ${\cal O}\left(10\,{\rm mK}\right)$, is much smaller than other typical energy scales of the system. The $\pi$-electrons graphene bandwidth is ${\cal O}\left({\rm eV}\right)$, the interlayer potential difference due to the displacement field (at the relevant parameter regime where superconductivity is observed) is ${\cal O}\left( 50\,{\rm mev}\right)$, and the distance between the vHS  and the top of the valance band, due to the trigonal warping [see Eq.~(\ref{eq:trigonalwarping})]  band is ${\cal O} \left(10 \,{\rm mev}\right)$~\cite{ZhouYoungBLGZeeman}.   

It is thus instructive to examine the expression for weak coupling superconducting critical temperature,
$T_c \sim \omega_c\exp\left(-\frac{1}{\tilde{g}\bar{{\cal N}}}\right)$, with the effective pairing interaction $\tilde{g}$, the pairing interaction cutoff $\omega_c$, and the Fermi level DOS $\bar{{\cal N}}$.
The dimensionless coupling constant is assumed to be rather small, $\tilde{g}\bar{{\cal N}}\ll 1$.

The important observation is that the critical temperature is extremely sensitive to slight changes in the Fermi level DOS in this case.
Quantitatively, one may relate the change in critical temperature $\delta T_c$, to a DOS variation $\delta\Bar{\cal N}$, 
\begin{equation}
    \frac{\delta T_c}{T_c} = \frac{1}{\tilde{g}\bar{{\cal N}}} \times \frac{\delta\Bar{\cal N}}{\Bar{\cal N}}.\label{eq:weakcouplinglever}
\end{equation}
Thus, in a weak-coupling scenario there is a huge ``lever factor'' converting DOS changes into modification of $T_c$.

As a consequence of the above considerations, blockading the high DOS regions of individual flavor fillings can catastrophically weaken superconductivity.
Conversely, relief of the Stoner blockade, even by a modest amount, may produce more robust superconductivity with higher $T_c$.
We now move on to discuss a natural way to lift the blockade -- via introducing a Zeeman term. 

\subsection{Softening the phase transitions}\label{sec:softeningtransitionsmagnetization}

Let us examine the width of the Stoner blockaded region $\Delta \nu_{b}$ more carefully. 
To that end, we introduce a simple model for the free energy exhibiting a first-order phase transition and a jump in the densities and magnetization.
Approximating the vHS as symmetric around the singular filling $n_{\rm vHS}$, we can relate the blockade to a \textit{first-order jump in magnetization} $\Delta m$ (Appendix~\ref{app:phenophasetransitionappendix}),
\begin{equation}
    \Delta \nu_b \approx \Delta m -\frac{1}{2} \left|n_{\rm vHS}-n^c\right|,\label{eq:blocakdemagnetization}
\end{equation}
with $n^c$ the density at which the phase transition spontaneously occurs, and we defined the magnetization $m\equiv \sum_{\tau,s} \sigma_z^{ss}\nu_{\tau s}$.
The upshot of the crude estimate in the expression~\eqref{eq:blocakdemagnetization} is that reducing the first-order magnetization jump immediately shrinks the blockaded region.

It is well-known that a spontaneous first-order transition is softened by a perturbation that couples linearly to the order parameter.
In the case of spin magnetization, this is clearly just a Zeeman magnetic field.
Consider the following simple free-energy density, expanded around the phase transition point,
\begin{equation}\label{eq:phenofreeenergymagnetization}
    f\left(m\right) = f_0 + \alpha m^2
    -\frac{1}{2}\beta m^4
    +\frac{1}{3}\gamma m^6
    -Bm.
\end{equation}
For simplicity, as we are only interested in the qualitative properties of the phase transition, in Eq.~\eqref{eq:phenofreeenergymagnetization} $m$ is the dimensionless order parameter (magnetization), $\alpha,\beta,\gamma>0$ have units of energy density, and $B$ is the Zeeman-like energy density. Notice $\alpha$ is the parameter that controls the transition (in our case, the relevant parameter is the electron density).
In terms of the above parameters, the $B=0$ transition, where the minimum of $f$ is at $m\neq 0$, occurs at $\alpha_c = \frac{3}{16}\frac{\beta^2}{\gamma}$, and the magnetization jumps by a magnitude $\Delta m^0 = \sqrt{\frac{3\beta}{4\gamma}}$.
By calculating the magnetic susceptibility $dm/dB$ on both sides of the transition, we find the small field dependence of this jump (Appendix~\ref{app:phenophasetransitionappendix}),
\begin{equation}\label{eq:magnetizationjumpphenomenological}
    \Delta m \approx \Delta m^0 - 2 \frac{\gamma}{\beta^2}B.
\end{equation}
One thus recovers the expected effect: a finite magnetic field significantly softens the first-order phase transition.

\begin{figure}
\begin{centering}
\includegraphics[scale=0.5,viewport=60bp 220bp 450bp 780bp]{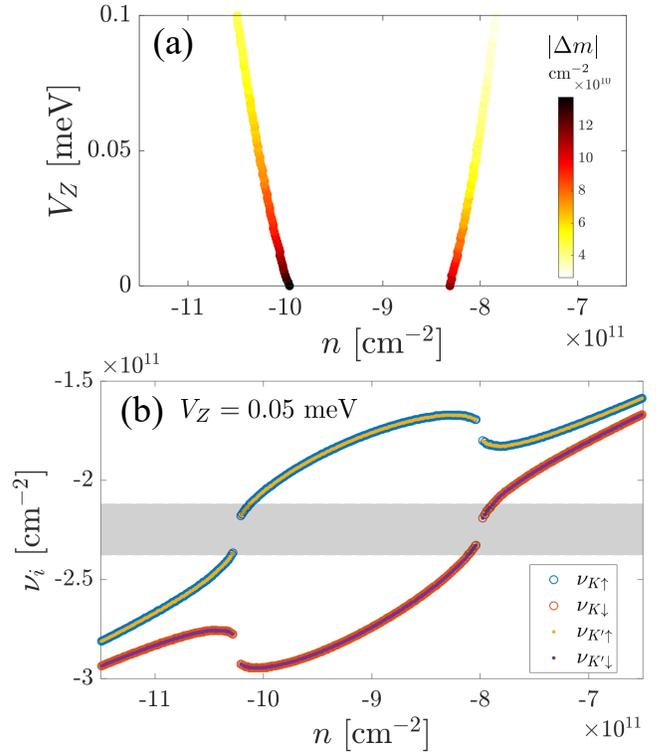}
\par\end{centering}
\caption{ \label{fig:softenedtransition}
(a) 
The magnitude of the discontinuous magnetization jump at the phase transition points, as a function of Zeeman coupling [Eq.~\eqref{eq:Zeemanterm}].
As expected from a first-order magnetization transition, the jump softens with an increase in the magnetic field.
Here, the magnetization is defined $m\equiv \sum_{\tau,s} \sigma_z^{ss}\nu_{\tau s}$.
(b)
Flavor resolved densities $\nu_i$ as a function of total electron filling, with $V_Z=0.05$ meV.
The gray rectangle marks the forbidden range of flavor density when $V_Z=0$ (see Fig.~\ref{fig:StonerBlockade}b).
Notice that in the vicinity of the transition, some flavors occupy a previously-forbidden region.
Other than $V_Z\neq 0$, the parameters used in this Figure are identical to the ones in Fig.~\ref{fig:StonerBlockade}b.
}
\end{figure}

Let us now turn to include this effect explicitly within our model by  introducing the Zeeman coupling
\begin{equation}
    {\cal H}_{\rm SB}^{\rm Zeeman} = -V_Z
    \sum_{\mathbf k}\Psi^\dagger_ {\mathbf {k}}s_z\Psi_ {\mathbf {k}},\label{eq:Zeemanterm}
\end{equation}
which explicitly breaks the spin $SU\left(2\right)$ symmetry of $H_0+H_{\rm int}$.
Repeating our variational analysis with finite $V_Z$ we find precisely the expected behavior from the above simplified considerations.
Namely, the jump in magnetization gradually decreases on both sides of the transition, as illustrated in Fig.~\ref{fig:softenedtransition}a.

In Fig.~\ref{fig:softenedtransition}b we demonstrate the effect of finite Zeeman coupling on the so-called blockade.
The flavor-resolved densities now encroach into the previously forbidden territory in the $V_Z=0$ case.
Thus, the normal state Fermi level DOS may become higher with applied in-plane magnetic fields.

\section{Superconductivity}\label{sec:superconductivitycalculations}
Having established the relevant phenomenon naturally arising in the non-superconducting normal state of BLG, we explore its effects on superconductivity in this system.
Our starting point for this discussion will be the result of the variational Hartree-Fock approach.
For simplicity, we assume superconductivity emerges as a consequence of pairing between same-spin electrons from opposite valleys, corresponding to the flavors which are in closer vicinity to the vHS.
We therefore disregard the other two flavors whose Fermi energy resides far away from the vHS.
We assume intervalley electron pairing, as finite-momentum pairing is generically considered less favorable due to its sensitivity to disorder effects.

\subsection{Tolmachev-Anderson-Morel approach}\label{sec:andersonmorel}
Projecting on to the valence bands of $H_0$ We consider the action
\begin{align}\label{eq:superconductivityaction}
    {\cal S} &= \sum _{n,{\mathbf k},\tau}\left(\xi_{{\mathbf k},\tau}-i\omega_n\right) \Bar{c}_{n{\mathbf k}\tau}c_{n{\mathbf k}\tau}\nonumber\\
    &+\sum_{n,m,\ell,\mathbf{k},\mathbf{k'},\mathbf{q},\tau,\tau'} V_{\mathbf q} 
    \Bar{c}_{n+\ell,\mathbf{k+q}\tau} c_{n\mathbf{k}\tau} \Bar{c}_{m-\ell,\mathbf{k'-q}\tau'} c_{m\mathbf{k'}\tau'},
\end{align}
where $c_{n{\mathbf k}\tau}$ is a fermionic Grassman variable corresponding to a fermion with Matsubara frequency $\omega_n=\pi\left(2n+1\right)T$, momentum $\mathbf k$ at valley $\tau$, $\xi_{{\mathbf k},\tau}=\epsilon_{{\mathbf k}\tau}-\bar{\mu}$, $\epsilon_{{\mathbf k}\tau}$ is the electronic spectrum of the valley $\tau$ valence band, $\bar{\mu}$ is the Fermi energy of the relevant sector, and $V_{\mathbf q}$ is a generalized interaction projected onto the BLG valence bands.
We simplify the interaction term by replacing the general $V_{\mathbf q}$ with a single short-range term $V$, and decouple the interaction term in the Cooper channel (pairing between electrons of opposite momenta at opposite valley) via a Hubbard-Stratonovich transformation, such that the action reads
\begin{align}
    \tilde{{\cal S}}&= \sum _{n,{\mathbf k},\tau}\left(\xi_{{\mathbf k},\tau}-i\omega_n\right) \Bar{c}_{n{\mathbf k}\tau}c_{n{\mathbf k}\tau}
    +\frac{1}{V}\bar{\Delta}\Delta
    \nonumber\\
    &+i\sqrt{\frac{T}{\Omega}}\sum_{n,{\mathbf k},\tau}\left(
    \Delta \Bar{c}_{n{\mathbf k}\tau}\Bar{c}_{-n,-{\mathbf k},-\tau}
    +\Bar{\Delta} {c}_{-n,-{\mathbf k},-\tau}{c}_{n{\mathbf k}\tau}
    \right),
\end{align}
where $\Delta$ is the superconducting order parameter.

Our analysis thus proceeds in two steps. 
We first assume an upper energy cutoff on the action $\Lambda_0$, at which the interaction is repulsive.
The initial value of $V$ will be determined by the screened Coulomb interaction, $V_{\mathbf q}\approx 2\pi e^2 /\left(\epsilon_r \left|\mathbf q\right|\right)$ ($\epsilon_r$ is the dielectric constant).

Considering only low-momentum scattering (intervalley scattering is largely suppressed), the relevant momenta are of order $k_F\sim \sqrt{\left|n\right|/4}$ (accounting for the four flavors), which in our regime of interest is $\sim 1/20$ nm$^{-1}$.
However, the Thomas-Fermi momentum $q_{TF}=2\pi e^2{\cal N}\left(\bar{\mu}\right)/\epsilon_r$ is of order ${\cal O} \left(1\,{\rm nm}^{-1}\right)$ (considering $\epsilon_r=4$ for hBN and the vicinity of the vHS), i.e., much larger than $k_F$.
Physically, this indicates that the combination of large Fermi energy DOS with low electron density means the Coulomb repulsion is quite efficiently screened.
We will thus replace $V_{\mathbf q}\approx{\cal N}^{-1}\left(\bar{\mu}\right)$ henceforth. 
Similar considerations were discussed in Refs.~\cite{DasSarmaBLGphonons,LianBioBernevigPRLphonons}.

We will also include the effects of the Hund's interaction, which is attractive in the spin-polarized Cooper channel, such that the initial interaction is
\begin{equation}
    V\left(\Lambda_0\right) = {\cal N}^{-1}\left(\Bar{\mu}\right) - J.
\end{equation}
We note that $V\left(\Lambda_0\right)$ is still positive, since we take the subleading interaction term $J$ to be much smaller than the dominant Coulomb repulsion energy scale.

In the first step, we integrate out high energy electrons down to $\omega^*$, the scale at which retarded attractive interactions come in.
In the case of acoustic phonon mediated attraction, this would be the Debye frequency.
Keeping only the leading term in $\Delta$, since we are interested only in the vicinity of the superconducting transition, the effective interaction at this point is somewhat reduced,
\begin{equation}
    V\left(\omega^*\right) ^{-1}=
    V\left(\Lambda_0\right)^{-1}
    +
    \frac{1}{2\Omega}\sum^*
    \frac{1+{\rm sgn}\left(\xi_{\mathbf{k},\tau}\xi_{-\mathbf{k},-\tau}\right)}{\left|\xi_{\mathbf{k},\tau}\right|+\left|\xi_{-\mathbf{k},-\tau}\right|}
    ,\label{eq:renormalizedVstar}
\end{equation}
where the sum $\sum^*$ is over energies $\omega^*<\left|\xi_{{\mathbf{k},\tau}}\right|\leq\Lambda_0$.
We have taken the limit $T\to 0$ in the above expression, since the energies integrated over are assumed to be much higher than the temperature.
Notice that we generally allow $\xi_{\mathbf{k},\tau}\neq\xi_{-\mathbf{k},-\tau}$, which is excluded by $H_0$, but will be made possible by orbital effects of the magnetic field.

In the next step, we introduce the attraction $g$ at the scale $\omega^*$, and calculate the vertex function $\chi_{\rm SC}$ by integrating out the remaining electrons (assuming $\left|g\right|>V\left(\omega^*\right)$),
\begin{multline}
    \chi_{\rm SC}^{-1} = \left[\left|g\right|-V\left(\omega^*\right)\right]^{-1}\\
    -\frac{1}{\Omega}\sum_{\left|\xi_{{\mathbf k},\tau}\right|\leq\omega^*}
    \frac{1-f\left(\xi_{\mathbf{k},\tau}\right)-f\left(\xi_{-\mathbf{k},-\tau}\right)}{\xi_{\mathbf{k},\tau}+\xi_{-\mathbf{k},-\tau}},\label{eq:superconductingsusceptibility}
\end{multline}
where $f\left(x\right)=1/\left(1+e^{x/T}\right)$.
We will extract $T_c$ as the temperature at which the vertex function diverges, i.e.,
\begin{equation}
    \chi_{\rm SC}^{-1}\left(T_c\right)=0.\label{eq:susceptibilitydiverges}
\end{equation}

\begin{figure*}
\begin{centering}
\includegraphics[scale=0.45,viewport=40bp 00bp 1500bp 300bp]{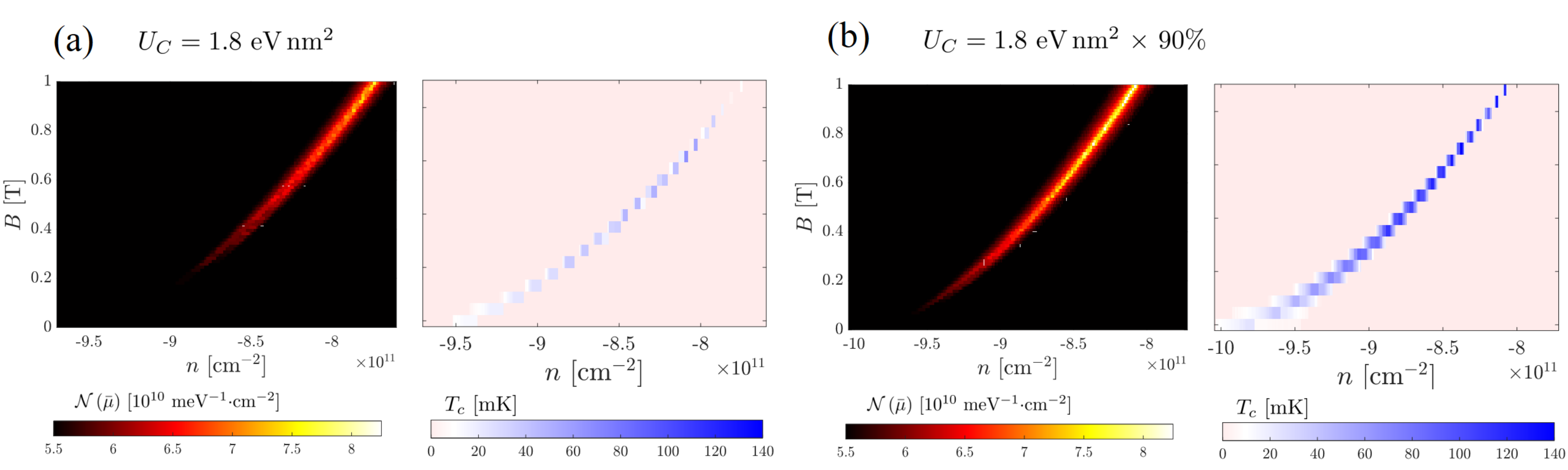}
\par\end{centering}
\caption{ \label{fig:megaTcfigure}
Superconductivity near the ferromagnetic phase transition boundary.
Left panels:
The DOS at the Fermi level in the superconducting sector as a function of density and in-plane magnetic field.
Right panels:
The corresponding superconducting transition temperature, calculated by the methods of Sec.~\ref{sec:superconductivitycalculations}.
Both features follow closely the phase transition line, as one would expect within the Stoner blockade mechanism.
(a)
Calculations with Coulomb repulsion parameter $U_C=1.8$ eV$\,$nm$^2$.
(b)
Same as (a), with $U_C$ reduced by 10\%.
Notice the colorscales are identical for panels (a) and (b), emphasizing the immense potential impact of slight modifications of the Coulomb interaction strength.
For example, the maximal $T_c$ increased by a factor of $\sim 2.5$ (66 mK to 158 mK) after the the 10\% reduction in $U_C$.
Other parameters used: $U=60$ meV, $J=0.25$ eV$\,$nm$^2$, $\Lambda_0=25$ meV, $\omega_c=0.6$ meV,  and $g=0.63$ eV$\,$nm$^2$ (the last three parameters are defined in Sec.~\ref{sec:andersonmorel}).}
\end{figure*}

We note that in order to obtain a finite magnetic field threshold for superconductivity, one needs (at some particular filling) the expression $\left|g\right|-V\left(\omega^*\right)$ to be negative at zero magnetic field, and to flip sign at the threshold field value.
Since the Zeeman term brings regions with larger DOS closer to the Fermi surface (i.e., decrease their relevant $\xi_{{\mathbf k},\tau}$), $V\left(\omega^*\right)$ decreases with increased magnetic field at the relevant fillings [as implied by Eq.~\eqref{eq:renormalizedVstar}], enabling the threshold effect at small enough values of $\left|g\right|$.

\subsection{Orbital magnetic field effect}\label{sec:orbitaleffects}
When an in-plane magnetic field is applied to a BLG device, the Zeeman term coupling to to electron spins is not the only perturbation to the Hamiltonian.
As a finite flux is penetrating the space between the graphene layers, one should modify $H_0\to H_0 + \sum_{\mathbf k}\Psi^\dagger_{\mathbf k}h_{\rm orb}\Psi_{\mathbf k}$, with~\cite{BLGmagneticfieldorbital}
\begin{equation}
    h_{\rm orb} = \frac{2v^2}{\gamma_1}
    \left(
    {\mathbf b} \times {\mathbf k}
    \right)_z
    \left( \frac{v_4}{v}\sigma_z + 2\frac{U}{\gamma_1}
    \right), \label{eq:orbitalmagneticfield}
\end{equation}
where ${\mathbf b}=e{\mathbf B}d/2$, $\mathbf B$ is the in-plane magnetic field, $d$ is the interlayer separation, and we only consider leading-order terms in $v_4/v$ and $U/\gamma_1$. For definition of the different parameters, see Eq.~\eqref{eq:detailsofsinglebandterms}.
We have verified numerically  that $h_{\rm orb}$ has a negligible effect on the phase transitions studied in Sec.~\ref{sec:normalstatecascade} for experimentally relevant magnetic fields of order $\cal O$ $\left(1\,{\rm T}\right)$ or less.

Notice that $h_{\rm orb}$ is odd in momentum $\mathbf k$ and even with respect to valley.
The relevant effect of this term regarding superconductivity is to make $\epsilon_{{\mathbf k},\tau}\neq \epsilon_{-{\mathbf k},-\tau}$, resulting in a non-negligible pair-breaking effect.
Although the orbital energy is rather small compared to the Zeeman energy associated with the magnetic field (due to small layer separation $d$, and relevant Fermi momenta), it becomes important compared to the tiny superconducting $T_c$s which are presumably realized in experimental devices.
As can be seen in Fig.~\ref{fig:megaTcfigure}, this leads to a narrowing of the superconducting region with increased $B$, whereas the pure Zeeman effect would not lead to such an effect.
The latter can be understood from the fact that the Fermi level DOS grows monotonically with magnetic field, just as the Stoner blockade picture would imply.

An important consequence of the Stoner-blockaded superconductivity at zero field, the mechanism that we propose here, is an \textit{extraordinary sensitivity to Coulomb repulsion strength}.
Let us compare panels (a) and (b) in Fig.~\ref{fig:megaTcfigure}, where in the latter we slightly reduce the interaction parameter $U_C$ by a mere 10\% compared to the former.
As one might expect from the discussion in Sec.~\ref{sec:stonerblockade}, the Fermi levels in the reduced-repulsion-strength scenario may come much closer tot he vicinity of the vHS, significantly increasing the Fermi level DOS ${\cal N}\left(\bar{\mu}\right)$.

In turn, thanks to an effective weak-pairing lever factor [along the lines of Eq.~\eqref{eq:weakcouplinglever}], this gives rise to enhancement of superconductivity.
Both the superconducting transition temperature and the regions where superconductivity is stabilized are enhanced.
We stress that change here was made to $U_C$ alone, which determines the variational ground state.
The initial coupling in the Cooper channel, $V\left(\Lambda_0 \right)$ remains unaltered for panels (a)--(b). 
Thus, the effect we demonstrate in Fig.~\ref{fig:megaTcfigure} is not due to introducing additional attraction in the superconducting channel, but rather due to \textit{modification of the normal state properties}.

\section{Refinements}\label{sec:refinements}
\subsection{Ising spin-orbit coupling}\label{sec:isisngspinorbitisoc}
Inspired by the experiment in Ref.~\cite{ZhangBLGSOC}, we consider replacing the Zeeman term by a substrate-induced Ising spin-orbit coupling (ISOC), i.e.,
\begin{equation}
    {\cal H}_{\rm SB}^{\rm ISOC} = -\lambda_{\rm ISOC}
    \sum_{\mathbf k}\Psi^\dagger_ {\mathbf {k}}s_z\tau_z\Psi_ {\mathbf {k}},\label{eq:lambdaISOC}
\end{equation}
which promotes so-called spin-valley locking, with the spin in the out-of-plane direction.
The Stoner blockade mechanism explored in this work may also help explain the findings of Ref.~\cite{ZhangBLGSOC} regarding the stabilization of superconductivity in BLG when an ISOC-inducing substrate (e.g., $\rm WSe_2$) is included.
To demonstrate this, it is instructive to consider the following two limits.

\begin{figure}
\begin{centering}
\includegraphics[scale=0.48,viewport=120bp 480bp 450bp 780bp]{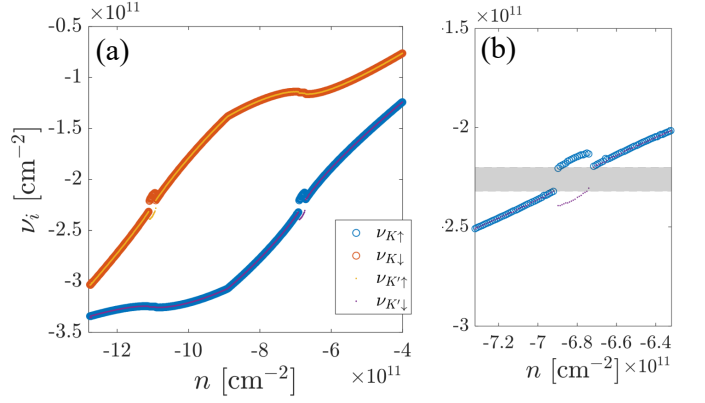}
\par\end{centering}
\caption{ \label{fig:ISOCfigure}
(a) 
Flavor resolved densities $\nu_i$ (calculated by the variational Hartree-Fock method) in the presence of strong ISOC, $\lambda_{\rm ISOC}=0.7$ meV, consistent with Ref.~\cite{ZhangBLGSOC}.
Here, $U_C=1.8$~eV$\,$nm$^2$, $J=-0.1$~eV$\,$nm$^2$.
(b)
Zoom-in on one of the blockaded region close to the van-Hove filling.
The blockaded region, where the intra-spin-valley sector spontaneously polarizes, thus suppressing intervalley pairing, is demarcated by a gray rectangle.
Notice the y-axis scale is the same as Fig.~\ref{fig:StonerBlockade}b, showing the blockaded region is significantly smaller due to the large ISOC.
}
\end{figure}

\begin{enumerate}

    \item \textit{Opposite sign (antiferromagnetic) Hund's interaction.--}
    As we demonstrate explicitly in Appendix~\ref{app:HartreeFock}, flipping the sign of the intervalley Hund's term $J\to -J$, exactly maps the scenario of spontaneous spin-polarization  ($\left\langle s_z\right\rangle_{\rm H.F.}\neq0$) in the presence of a Zeeman term, to polarization of different spin-valley locked sectors  ($\left\langle s_z\tau_z \right\rangle_{\rm H.F.}\neq0$) in the presence of ISOC.
    Whereas this limit is quite extreme, a plausible mechanism for the sign change of this term in the presence of a substrate is discussed in Appendix~\ref{app:hundsderivation}.

    Thus, a moderate amount of ISOC, $\lambda_{\rm ISOC}\sim{\cal O}\left(1\,{\rm meV}\right)$ as measured in experiments, acts as an effective magnetic field of the order of 10 Tesla in the context of the Stoner blockade (although without the adverse orbital effects).
    One would thus expect Stoner-blockaded superconductivity to be much stronger in this scenario, as compared to the Zeeman-triggered one.
    This is of course entirely consistent with experimental results thus far~\cite{ZhangBLGSOC,BLGyoungNadjperge}.

    \item \textit{$\lambda_{\rm ISOC}\to\infty$.--}
    If the ISOC overwhelms the other energy scales in the problem, one may consider the scenario where one spin-valley sector is inert as it occupies some remote low-DOS region, whereas the other two flavors may develop some valley polarization near the van-Hove filling, (see Fig.~\ref{fig:ISOCfigure}).
    This will thus suppress pairing between electrons in this sector, in what we will denote as ``mini-blocakde''.

    Comparing with the spontaneously developed spin polarization in the $\lambda_{\rm ISOC}=0$ case, one has half the DOS.
    Thus, whereas the former blockade is dominated by the interaction $U_C$, the mini-blockade is effectively controlled by $U_C/2$.
    As one gleans from Fig.~\ref{fig:StonerBlockade}c, this significantly reduces the blockaded region, hence stabilizing superconductivity.

\end{enumerate}

Whereas these two limits are probably far from exact in the experimental scenario, they help us make sense of ISOC-enhanced superconductivity in the context of the Stoner blockade mechanism presented here.

We provide a more quantitative example of the cascade of phase transitions in the presence of ISOC (comparable in magnitude to experiments) in Fig.~\ref{fig:ISOCfigure}.
For the most part, the ISOC splits the occupation the two spin-valley locked sectors by an amount which gradually increases as the two sectors Fermi energy approaches the vHS, as expected.
This effect is similar, up to a change of flavor labels, to the separation of spin-polarized sectors in the presence of a large Zeeman energy.
Sufficiently close to the vHS, we observe the intra-sector mini-blockade dominated by a smaller $\sim U_C/2$ repulsion.

Qualitatively, $\lambda_{\rm ISOC}$ should increase monotonically with the displacement field, up to some saturation field.
This is due to the spin-orbit term originating in proximity to a $\rm WSe_2$ layer, which is maximized when the valence band electron wavefunctions are entirely layer polarized ($\left\langle\sigma_z\right\rangle\approx 1$ in our notation).
Therefore, our theory predicts the superconducting region to expand and $T_c$ to increase with the growing displacement field, consistent with the experimental scenario~\cite{ZhangBLGSOC,BLGyoungNadjperge}.

\subsection{Insufficiency of an Anderson's theorem}\label{sec:andersonstheoremdisorder}

In this Section, we argue that the superconductivity described here is remarkably delicate and disorder-sensitive, despite the simple pairing channel we consider is node-less, and even in the presence of so-called ``protection'' by Anderson's theorem~\cite{ANDERSONtheorem}. A disorder that leads to density fluctuations will blunt the vHS, reduce the DOS at its vicinity, and will cause a decrease in the superconducting critical temperature.

Within our spin-polarized valley-degenerate subspace, and neglecting orbital magnetic field effects, we write the mean-field superconducting Hamiltonian as 
\begin{equation}
    H_{\rm SC} = \sum_{\mathbf k}
    C^\dagger_{\mathbf k}
    \left(
    \Bar{\xi}_{\mathbf k}\nu_z + \delta\xi_{\mathbf k}\tau_z +\Delta \tau_x\nu_x
    \right)
    C_{\mathbf k},\label{eq:meanfieldsuperconductingbcs}
\end{equation}
with the Nambu spinor 
$C_{\mathbf k}=\left(c_{{\mathbf k},+},c_{{\mathbf k},-},c^\dagger_{-{\mathbf k},+},c^\dagger_{-{\mathbf k},-}\right)^T$,
and the Pauli matrices $\tau_i$ and $\nu_i$ operating on valley and particle-hole space, respectively.
We also defined
$\Bar{\xi}_{\mathbf k}=\left(\xi_{{\mathbf k},+}+\xi_{{\mathbf k},-}\right)/2$ and $ \delta\xi_{\mathbf k}=\left(\xi_{{\mathbf k},+}-\xi_{{\mathbf k},-}\right)/2$.

It is instructive to apply a unitary transformation (along the lines of Ref.~\cite{matbgKIVCanderson}) $C_{\mathbf k}\to {\cal U} C_{\mathbf k}$, with ${\cal U} = \frac{1+\nu_z}{2}+ \frac{1-\nu_z}{2}\tau_x$.
The transformed Hamiltonian at momentum $\mathbf k$ is
$h_{\mathbf k} = 
    \left(\Bar{\xi}_{\mathbf k}+\delta\xi_{\mathbf k}\tau_z\right)\nu_z   +\Delta \nu_x$. 
This Hamiltonian has intrinsic particle-hole symmetry, i.e., it anti-commutes with the unitary 
${\cal P}=\nu_y \tau_y {\mathcal{K}}$ ($\cal K$ is the complex conjugation operator).
Notice that although the phase we consider is spin-polarized, the Hamiltonian still possesses a residual spinless time-reversal symmetry, 
${\cal T}=\tau_x {\mathcal{K}}$.

The key observation regarding disorder here, is that any perturbation to the normal state which is $\cal T$-symmetric (and also adheres to $\cal P$ by construction) is proportional to $\nu_z$.
Thus, such perturbations anti-commute with the superconducting order parameter $\Delta\nu_x$.
In this scenario, it has been shown~\cite{ANDERSONtheorem,abrikosovandersonthorem,maki1969gapless} that the only change to the self-consistent superconducting gap equation is replacement of the DOS of the pristine Hamiltonian $H_0$, by that of the perturbed normal-state.
For example, Eq.~\eqref{eq:superconductingsusceptibility} will be modified as (notice once more we do not consider the orbital effect of the magnetic field in this section),
\begin{equation}
     \chi_{\rm SC}^{-1} = \left[\left|g\right|-V\left(\omega^*\right)\right]^{-1}
    -2\int_0^{\omega^*} d\xi \tilde{\cal N}\left(\xi\right)
    \frac{\tanh\left(\frac{\xi}{2T}\right)}{\xi},\label{eq:superconductingsusceptibilityAnderson}
\end{equation}
where $\tilde{\cal N}$ is the DOS in the presence of disorder.

In graphene, the relevant sources of disorder are ripples~\cite{ripplesIshigami,ripplesStolyarova}, charge impurities~\cite{ImpuritiesIshigami}, and strain variations~\cite{StrainDisorderMapping}. 
It has been argued that strain disorder, which acts as a random gauge field~\cite{RMPgraphene}, is the dominant type of disorder in state-of-the-art graphene devices~\cite{grapheneStrainDisorderPRX}, and plays an important role in twisted graphene multilayers~\cite{straindisordertbg}.
In any case, these all inherently preserve the spinless time-reversal symmetry, so that one may apply Anderson's theorem to the superconductivity discussed above.

However, \textit{this does not necessarily mean superconductivity persists in the presence of such disorder}.
To simplify the remaining discussion and illustrate our point, we ascribe a single parameter to describe the strength of disorder in the system -- the charge inhomogeneity $\delta n$.
This quantity is usually extracted as roughly the width of the resistance peak of a graphene device at charge neutrality and zero displacement field~\cite{chargeinhomegeneitypeakwidth}.
It has been previously shown to be directly related to the mobility in monolayer graphene and BLG devices~\cite{grapheneStrainDisorderPRX}, and thus will provide a useful metric for our discussion.

\begin{figure}
\begin{centering}
\includegraphics[scale=0.52,viewport=90bp 480bp 450bp 780bp]{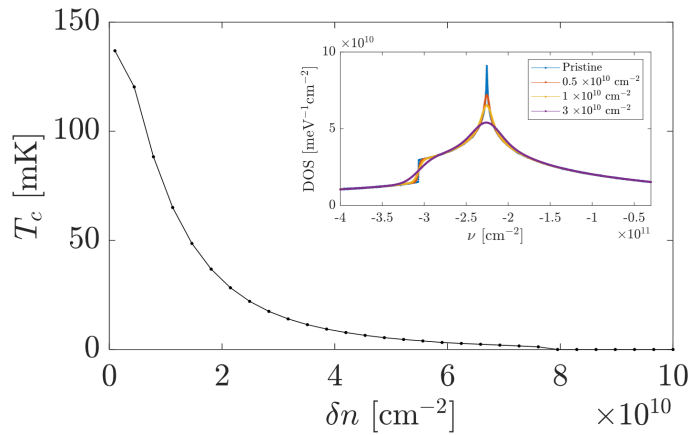}
\par\end{centering}
\caption{ \label{fig:disorderfigure}
Main panel: Superconducting $T_c$ as a function of charge inhomogeneity $\delta n$ induced by time-reversal symmetric disorder.
The calculation was done for $U=60$ meV and $\Bar{\mu}=-57.85$ meV, with parameters $\Lambda_0=25$ meV, $\omega_c=0.6$ meV, $J=0.25$ eV$\,$nm$^2$, and $g=0.65$ eV$\,$nm$^2$.
Inset:
Comparison of the pristine DOS to the broadened DOS for several values of $\delta n$ (indicated by legend).
}
\end{figure}

We consider only the effect of DOS broadening brought on by inhomogeneity.
We thus broaden the computed DOS by convolution with a normal distribution with standard deviation $\sigma\approx \delta n/2.355$ (such that $\delta n$ is the full-width-at-half-maximum of the distribution).
Examples of the DOS broadening can be found in the inset to Fig.~\ref{fig:disorderfigure}.
One clearly sees that the immediate casualty of the broadening is the vHS, which loses much of its sharpness.
Recalling Eq.~\eqref{eq:weakcouplinglever}, this suppression of the DOS peak may have dire consequences for superconductivity in this system.

Let us now demonstrate this point.
We repeat our procedure of extracting the superconducting $T_c$ from Sec.~\ref{sec:superconductivitycalculations} as a function of disorder.
The effect on the critical temperature in the scenario where the Fermi energy is close to the vHS is illustrated in Fig.~\ref{fig:disorderfigure}.
Superconductivity, in this case, is quite delicate and sensitive to even a moderate amount of charge inhomogeneity.
As a consequence, the theory presented here predicts (or rather post-dicts) that only exceptionally high-quality devices are expected to display the unique phenomenon of delicate superconductivity triggered by a magnetic field.
These are devices where the charge inhomogeneity is of order $10^{10}$ cm$^{-2}$ or lower.
This is precisely the order of magnitude of disorder in current state-of-the-art devices~\cite{Dean2010hBN,doi:10.1063/1.3665405hBN,Dean2011hBN}, providing a sensible explanation for the relative elusiveness of superconductivity in BLG devices.

\section{Conclusions}\label{sec:conclusions}

In this work, we have presented a theory of delicate superconductivity which is brought to light by either a finite magnetic field or spin-orbit coupling of the Ising type.
We dub the underlying mechanism at work here the Stoner blockade.
Namely, strong electronic correlation tends to cause spontaneous polarization and reconstruction of the Fermi surfaces, steering them away from the vicinity of the vHS and high DOS fillings.
Coupled with weak-enough electron pairing interactions, this scenario is devastating for superconductivity, which would have otherwise persisted in the non-reconstructed case (i.e., zero Coulomb repulsion).

However, we have shown that external perturbations, e.g., in-plane magnetic fields, may significantly alleviate the blockade under the right circumstances.
Moreover, the highest Fermi-level DOS, and thus strongest superconductivity, would be expected to occur in the vicinity of the symmetry-braking phase transition.
Such circumstances are consistent with the experimental observations in BLG~\cite{ZhouYoungBLGZeeman,ZhangBLGSOC,BLGyoungNadjperge}.
It thus becomes entirely reasonable to have a scenario where superconductivity is absent (or too weak to detect) unless a strong enough magnetic field is applied, in which case superconductivity stabilizes on the magnetic phase transition line. 
This is precisely the previously-enigmatic phenomenology of BLG.

The mechanism depicted here gives rise to a clear and distinct prediction that may be tested experimentally.
Namely, a small modification of the electron-electron repulsion strength, achieved by, e.g., changing the distance of the BLG to the nearby metallic gates, is expected to have outsize effects on superconductivity, as illustrated in Fig.~\ref{fig:megaTcfigure}.
A small reduction of the repulsion strength would significantly boost superconductivity.
By virtue of allowing the Fermi energy to come closer to the vHS, the reduced repulsion will lead to an appreciable increase in $T_c$, and to lowering of the critical Zeeman energy required for superconductivity -- up to a point where superconductivity may be detected without a magnetic field at all.
This prediction is in contrast to previous works discussing BLG superconductivity, where reducing the distance to a metallic gate either increases $T_c$, but very weakly~\cite{DasSarmaBLGphonons}, or has the precise opposite effect of suppressing superconductivity (as the pairing is presumably mediated by the same Coulomb repulsion)~\cite{ChubukovLevitovAlternative,KWANBLG,AlejandroPacoBLG}.
The prediction we make here is enabled by the identification of the underlying BLG normal-state, which highly depends on electron-electron interactions, as the culprit of the observed anomalous magnetic-field-triggered superconductivity.

Our work highlights the universality of Stoner blockaded superconductivity in BLG.
We have demonstrated that in-plane magnetic field or Ising SOC perturbations to the BLG Hamiltonian are on equal footing in terms of bypassing the blockade and revealing a superconducting phase.
These two types of perturbations differ, though in detail.
An in-plane magnetic field has a non-negligible orbital effect on superconductivity.
This is despite the tiny intervalley pair breaking effect, as it is still sizable compared to the small superconducting transition temperature.
As we have shown, the presence of the secondary Hund's-type interaction also introduces subtle differences between the two $SU\left(2\right)$ symmetry-breaking perturbations, which will manifest through subtle details in the flavor-resolved Fermi surface structure in the normal state.
Moreover, the induction of Ising SOC in BLG, which requires close proximity to a $\rm WSe_2$, may further modify the Hamiltonian in important ways and depend on various details of the stacking itself~\cite{dassarmaBLGsUBSTARTEiNDUCEDEFFECTS}.
These are expected to bear an impact on the phase diagram, which we leave to future investigations.

Several noteworthy issues related to the experimental phenomenology of intrinsic superconductivity in BLG, are not addressed by the unusual mechanism presented here.
The precise nature of the electron pairing glue, be it phonons~\cite{DasSarmaBLGphonons,DasSarmaPhonosGraphene} or a Kohn-Luttinger-like mechanism~\cite{ChubukovLevitovAlternative,BErgHolderPrbkohnluttingergraphene,KWANBLG}, is intentionally kept ambiguous.
Various possibilities may comfortably fit within our framework, which requires only that the pairing glue itself is weak enough such that there is an appreciable lever effect with regards to small Fermi-level DOS modifications [Eq.~\eqref{eq:weakcouplinglever}].
The related issue of an unconventional nodal pairing symmetry, e.g., p-wave (cf. Ref.~\cite{annularFermiSC}), is also unresolved.
For the sake of clarity, we just considered the simplest possible pairing channel, yet the conclusions drawn here should be generalized.

Experimentally, it is evident that superconductivity favors the vicinity of the phase transition boundary closer to charge neutrality over the boundary at higher hole-doping.
This is observed both in Refs.~\cite{ZhouYoungBLGZeeman,ZhangBLGSOC}, where superconductivity appears only there, and Ref.~\cite{BLGyoungNadjperge}, where the superconductor is far more robust closer to charge neutrality. In our theory,
a weak inherent asymmetry in the DOS around the vHS does exist, leading to a small asymmetry in the phase transition itself.
Notice, for example, Fig.~\ref{fig:softenedtransition}a depicting the magnitude of the magnetization jump -- it is somewhat smaller for the superconductivity-favoring region, consistent with the Stoner blockade picture.
In Appendix~\ref{app:secondphaseboundary} we show results for the vicinity of the second phase boundary as well.
Indeed, superconductivity is found to be much weaker near it, due to the subtle microscopic details of the band structure near that transition.
This is of course consistent with the findings in Ref.~\cite{ZhouYoungBLGZeeman}.
We note that the differences in $T_c$ that we observe in the vicinity of these two phase boundaries may become less significant when the pairing glue becomes somewhat stronger, or the Coulomb repulsion is modified.
We thus do not entirely rule out a scenario where small DOS effect we observe in our phenomenological Stoner blockade description are greatly enhanced by the nature of the pairing itself, its dependence on electron density, or the Fermi-surface topology~\cite{AlejandroPacoBLG,BErgHolderPrbkohnluttingergraphene}.

We finally comment on the zero-field normal state, which is observed to be more resistive near the magnetic phase transition, whose origin is not yet well-understood.
Our analysis does not exclude the possibility of a correlated insulator that onsets at a low enough temperature~\cite{ChubukovLevitovAlternative} or the emergence of an intervalley coherent spontaneous order~\cite{AshvinRTGIVC,ChatterjeeIVCRTG}.
We would like, however, to put forward another possibility, which is natural, given the mechanism explored here.
Namely, the formation of a micro-emulsion of the fully-symmetric and spin-polarized phases, which is argued to inevitably occur in the vicinity of a first-order phase transition of the kind discussed here~\cite{KivelsonSpivakFirstOrder}.
Since the two constituent phases have different densities, magnetizations, and Fermi-surface topology, it is reasonable to expect that domain walls should contribute to the overall resistivity.
In this scenario, the resistivity would peak near the phase transition as long as superconductivity has not emerged.
The qualitative and quantitative feasibility of this crudely-described mechanism require further investigation.

\begin{acknowledgments}
This project was partially supported by grants from the ERC under the European Union’s Horizon 2020 research and innovation programme (grant agreement LEGOTOP No 788715), the DFG CRC SFB/TRR183, the BSF and NSF (2018643), the ISF (1335/16), and the ISF Quantum Science and Technology (2074/19).
\end{acknowledgments}

\begin{widetext}
\begin{appendix}
\section{Variational Hartree-Fock calculation}\label{app:HartreeFock}

This appendix explains the details of the Hartree-Fock calculations we have performed.
We begin by computing the grand-potential associated with the normal-state Hamiltonian, Eq.~\eqref{eq:normalstateHamiltoniangeneral},  at a given chemical potential,
$\Phi=\left\langle H - \mu N_0\right\rangle_{\rm H.F.}$,
where $\left\langle\right\rangle_{\rm H.F.}$ denotes the expectation value calculated using the variational wavefunction appearing in Eq.~\eqref{eq:variationalwavefunctionpsi}, describing possible flavor symmetry breaking phases.
We define the flavor-resolved densities and kinetic energies, denoted by indices ($\tau,s$) for (valley, spin), as
\begin{equation}
    \nu_{\tau s} = \int_0^{\mu_{\tau s}} d\epsilon {\cal N} \left(\epsilon\right), \,\,\,\,\,
    {\cal E}_{\tau s} = \int_0^{\mu_{\tau s}} d\epsilon {\cal N} \left(\epsilon\right)\epsilon,
\end{equation}
where ${\cal N} \left(\epsilon\right)$ is the density of states per flavor obtained from Eq.~\eqref{eq:singleparticleh0}, and $\mu_{\tau s}$ are the variational chemical potentials [denoted by $\mu_i$ with a single flavor index in Eq.~\eqref{eq:variationalwavefunctionpsi}].

Combining the different ingredients of $H$, accounting for the possible Zeeman and Ising spin-orbit terms, a straightforward calculation allows one to obtain the grand potential density,
\begin{align}
    \frac{\Phi}{\Omega} &=\sum_{\tau s}\left[{\cal E}_{\tau s}+\left(-\mu+V_{Z}s_{z}^{ss}+\lambda_{{\rm ISOC}}\sigma_{z}^{ss}\tau_{z}^{\tau\tau}\right)\nu_{\tau s}\right] \nonumber\\
    &+\frac{1}{2}\sum_{\tau s\tau' s'}\nu_{\tau s}\left[U_{C}\left(1-\delta^{ss'}\delta^{\tau\tau'}\right)+\left(U_{V}-J
    \left(\delta^{ss'}-s_x^{ss'} \right)
    \right)\tau_{x}^{\tau\tau'}\right]\nu_{\tau's'}.
\end{align}
We now compare two scenarios of possible flavor symmetry breaking.

(i) \textit{Spin polarized (SP), valley degenerate, $\lambda_{\rm ISOC}=0$ --} In this case there are two distinct $\mu_{\tau s}$, one for each spin.
We denote 
$\nu_{\uparrow}\equiv \nu_{+,{\uparrow}}=\nu_{-,{\uparrow}}$,
$\nu_{\downarrow}\equiv \nu_{+,{\downarrow}}=\nu_{-,{\downarrow}}$,
${\cal E}_{\uparrow}\equiv {\cal E}_{+,{\uparrow}}={\cal E}_{-,{\uparrow}}$, and
${\cal E}_{\downarrow}\equiv {\cal E}_{+,{\downarrow}}={\cal E}_{-,{\downarrow}}$
so that the grand potential $\Phi_{\rm SP}$ is
\begin{align}
     \frac{\Phi_{\rm SP}}{\Omega} &=2\left({\cal E}_{\uparrow}+{\cal E}_{\downarrow}\right)-2\mu\left(\nu_{\uparrow}+\nu_{\downarrow}\right)+2V_{Z}\left(\nu_{\uparrow}-\nu_{\downarrow}\right)\nonumber\\
     &+\left(\nu_{\uparrow}+\nu_{\downarrow}\right)^{2}\left(\frac{3}{2}U_{C}+U_{V}\right)\nonumber\\
     &-\left(\nu_{\uparrow}-\nu_{\downarrow}\right)^{2}\left(\frac{U_{C}}{2}+J\right).\label{eq:HFappendixspinpolarized}
\end{align}

(ii) \textit{Spin-valley locked (SVL), $V_Z=0$ --}
Here, we denote
$\nu_{1}\equiv\nu_{+,\uparrow}=\nu_{-,\downarrow}$,
$\nu_{2}\equiv\nu_{+,\downarrow}=\nu_{-,\uparrow}$,
${\cal E}_{1}\equiv{\cal E}_{+,\uparrow}={\cal E}_{-,\downarrow}$,
${\cal E}_{2}\equiv{\cal E}_{+,\downarrow}={\cal E}_{-,\uparrow}$,
and we find $\Phi_{\rm SVL}$,
\begin{align}
     \frac{\Phi_{\rm SVL}}{\Omega} &=2\left({\cal E}_{1}+{\cal E}_{2}\right)-2\mu\left(\nu_{1}+\nu_{2}\right)+2\lambda_{{\rm ISOC}}\left(\nu_{1}-\nu_{2}\right)\nonumber\\
     &+\left(\nu_{1}+\nu_{2}\right)^{2}\left(\frac{3}{2}U_{C}+U_{V}\right)\nonumber\\
     &-\left(\nu_{1}-\nu_{2}\right)^{2}\left(\frac{U_{C}}{2}-J\right).\label{eq:HFappendixspinvalleylocked}
\end{align}
Notice that up to a change of labels of the different flavors, $\Phi_{SVL}$ is identical to $\Phi_{\rm SP}$ with the replacements $V_Z\to\lambda_{\rm ISOC}$ and $J\to -J$.

\section{First-order phase transition and the Stoner Blockade}\label{app:phenophasetransitionappendix}

Our purpose here is to relate the magnetization jump at the phase transition points to the Stoner-blockaded region of flavor-resolved densities.
For simplicity, we will assume that the transition is symmetric around the vHS density $n_{\rm vHS}$, such that the two critical densities $n^c_1$ and $n^c_2$ are equidistant from $n_{\rm vHS}$, and the magnetization jump $\Delta m$ is also identical at both transition points, see Fig.~\ref{fig:blocakdetomagnetizationapp}.
It thus becomes clear that the size of the blockade region is
\begin{equation}
    \Delta \nu_b = \left(\frac{1}{4} n^c_1 + \frac{\Delta m}{2}\right)
    -
    \left(\frac{1}{4} n^c_2 - \frac{\Delta m}{2}\right)
    =
    \Delta m + \frac{1}{2} \left|n_{\rm vHS} -n^c\right|,\label{eq:appblockademagnetizatio}
\end{equation}
where we have suppressed the number index of the critical point at the right hand side for simplicity, and recovered Eq.~\eqref{eq:blocakdemagnetization}.

\begin{figure}
\begin{centering}
\includegraphics[scale=0.6,viewport=90bp 420bp 450bp 780bp]{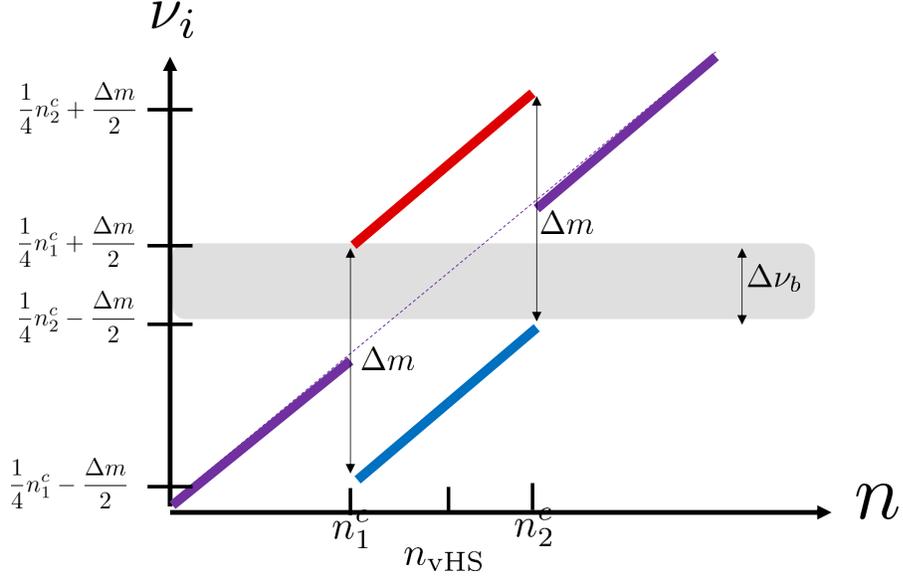}
\par\end{centering}
\caption{ \label{fig:blocakdetomagnetizationapp}
Relating the magnetization jump $\Delta m$ to the width of the Stoner blockaded region $\Delta \nu_b$.
The density $n$ ($x$-axis) controls the first-order magnetic transition and occurs at the critical densities $n^c_1$ and $n^c_2$.
For simplicity, we assume that the transition is symmetric around the vHS filling $n_{\rm vHS}$.
Below  $n^c_1$ and above $n^c_2$ the four flavors are equally occupied, $\nu_i = n/4$ (purple line).
At the transition points spontaneous magnetization of magnitude $\Delta m$ develops, so that opposite spin electrons have different densities (red and blue lines).
We assume the spin splitting is symmetric for simplicity.
By examining the region of flavor-resolved densities which is excluded by the spontaneous symmetry breaking, we arrive at the relation Eq.~\eqref{eq:appblockademagnetizatio}.
}
\end{figure}

For completeness, we detail the calculation of the magnetization jump at the transition.
Starting from the free-energy density in Eq.~\eqref{eq:phenofreeenergymagnetization} at $B=0$,
we find
\begin{equation}
    \frac{\partial f}{\partial m^2} = \alpha-\beta m^2+\gamma m^4.
\end{equation}
Combining the conditions for the phase transition
\begin{equation}
    \frac{\partial f}{\partial m^2} |_{\Delta m^0,\,\alpha_c}=0,
    \,\,\,\,\,
    f\left(\Delta m^0 \right)=f\left(0\right),
\end{equation}
one obtains $\left(\Delta m^0\right)^2=\frac{3\beta}{4\gamma}$, and $\alpha_c = \frac{3\beta^2}{16\gamma^2}$.

We now examine the magnetic susceptibility near the transition, and allow for finite infinitesimal $B$.
We obtain the saddle point equation $\partial f / \partial m=0$, and consider small variations of $m$ and $B$.
We find
\begin{equation}
    2 dm \left(\alpha - 3\beta m^2 +5\gamma m^4\right) = dB.
\end{equation}
Plugging in $\alpha=\alpha_c$, the susceptibilities on both sides of the transition are
\begin{equation}
    \frac{dm}{dB}_{m=0} = \frac{8\gamma^2}{3\beta^2},
    \,\,\,
    \frac{dm}{dB}_{m=\Delta m^0 } = \frac{2\gamma^2}{3\beta^2}.
\end{equation}
The difference in susceptibilities on both sides of the transition is responsible for the reduced magnetization jump, which is linear in $B$ (at small $B$), given by Eq.~\eqref{eq:magnetizationjumpphenomenological}.

\section{Modification of the Hund's coupling}\label{app:hundsderivation}
Here we demonstrate the mechanism by which the sign of the intervalley Hund's term may change in the presence of the substrate.
It has been shown in Ref.~\cite{dassarmaBLGsUBSTARTEiNDUCEDEFFECTS} that proximity to a $\rm WSe_2$ substrate tends to induce short-range \textit{attractive} interactions between electrons in the proximate layer.
Let us write a particular piece of this interaction, 
\begin{equation}
    H_{\rm inter} = \frac{1}{2\Omega}\sum_{\mathbf{k,k',q}}\sum_{s,s',\tau}
    \Tilde{U}
    A^\dagger_{{\tau}s\mathbf{k}}
    A_{\tau s \mathbf{k+q}}
    A^\dagger_{\bar\tau s'\mathbf{k'}}
    A_{\bar{\tau} s' \mathbf{k'-q}},\label{eq:appattractionintervalley}
\end{equation}
where $\Tilde{U}$ is the strength of the induced attraction (simplified to be extremely short-range),  and $A_{\tau s \mathbf{k}}$ annihilates an electron in layer $A$, valley $\tau$, spin $s$, and momentum $\mathbf k$.
Employing the Fierz identity $ \delta^{\alpha\beta}\delta^{\mu\nu} = 
    2\delta^{\alpha\nu}\delta^{\mu\beta} - 
    \mathbf{s}^{\alpha\beta}\cdot\mathbf{s}^{\mu\nu}$
with respect to the spin indices in Eq.~\eqref{eq:appattractionintervalley}, we may extract an intervalley Hund's term 
\begin{equation}
    H_{\rm Hund} = -\frac{1}{\Omega}\sum_{\mathbf{k,k',q}}
    \Tilde{U}
    \left(
    A^\dagger_{+,\alpha,\mathbf{k}}
    \mathbf{s}^{\alpha\beta}
    A_{+,\beta, \mathbf{k+q}}
    \right)
    \cdot
   \left(
    A^\dagger_{-,\mu,\mathbf{k'}}
    \mathbf{s}^{\mu\nu}
    A_{-,\nu, \mathbf{k'-q}}
    \right).\label{eq:appattractionhundsextratced}
\end{equation}
Crucially, the minus sign signals that for attraction, $\Tilde{U}<0$, the induced intervalley Hund's interaction is \textit{antiferromagnetic}, i.e., of the opposite sign of the presumed intrinsic ferromagnetic one in the absence of the $\rm WSe_2$ substrate.

Let us finally note that when projecting Eq.~\eqref{eq:appattractionhundsextratced} to the valence band electrons, one must consider momentum-dependent form factors.
However, in the regime of interest where superconductivity is observed, and where we perform our analysis, the applied vertical electric displacement field polarizes the valence bands electrons almost completely to the $A$ layer.
Thus, one expects these form factors may be fairly approximated by unity.

\section{Superconductivity near the lower-density phase boundary}\label{app:secondphaseboundary}
In the experiment, and within our phenomenological description of the Stoner blockade in BLG, there are two phase transitions separating the symmetry-broken phase from the flavor symmetric ones.
In the main text we have mainly focused on the transition which features superconductivity in Ref.~\cite{ZhouYoungBLGZeeman}.
Here, we provide the calculations that show, that under similar circumstances to the ones considered in the main text (Fig.~\ref{fig:megaTcfigure}a), superconductivity is much weaker or entirely absent in the lower-density phase boundary.

In the left panel of Fig.~\ref{fig:seconphaseboundary} the evolution of the Fermi-level DOS at the phase boundaries as a function of in-plane magnetic field is shown.
Both boundaries appear quite similar.
However, the left phase boundary has a slightly lower DOS, suggesting, it might be less susceptible towards developing superconductivity.
The right panel of Fig.~\ref{fig:seconphaseboundary} shows that this is in fact the case: superconductivity is much weaker at the lower-density phase boundary, and is completely washed away by the minuscule orbital effects at a much lower field.

\begin{figure}
\begin{centering}
\includegraphics[scale=0.5]{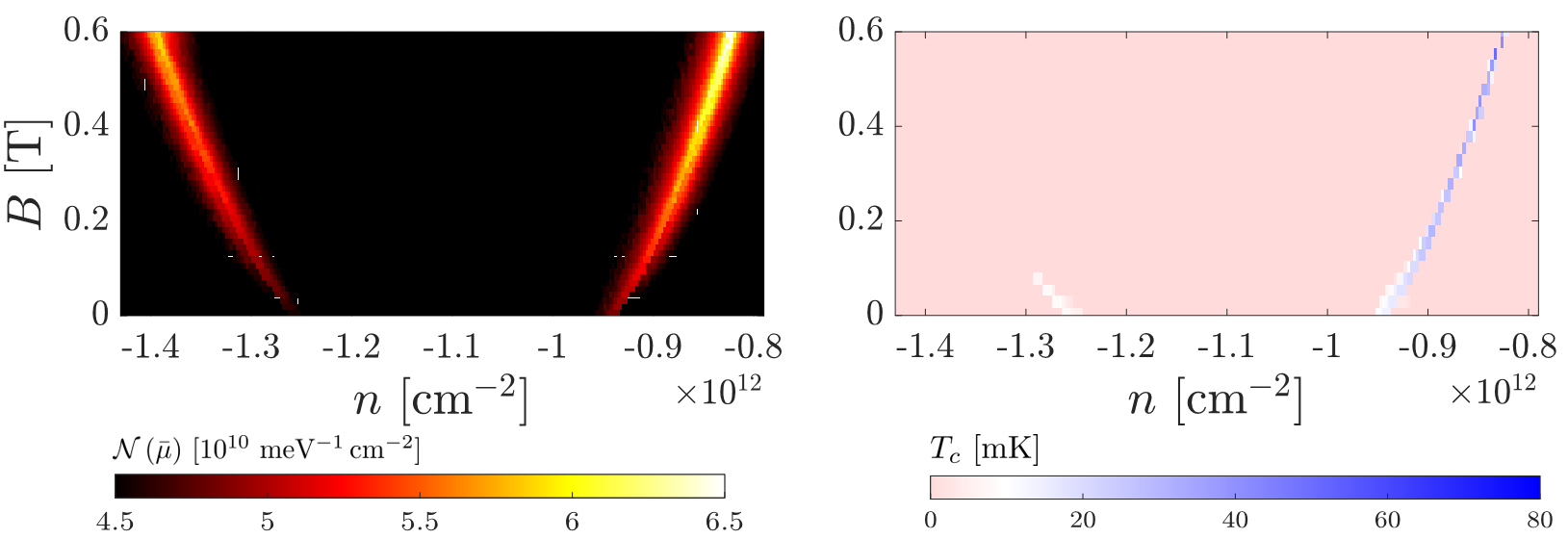}
\par\end{centering}
\caption{ \label{fig:seconphaseboundary}
Comparing the two ferromagnetic phase transition boundaries.
Left:
The DOS at the Fermi level in the superconducting sector as a function of density and in-plane magnetic field.
Right:
The corresponding superconducting transition temperature, calculated by the methods of Sec.~\ref{sec:superconductivitycalculations}.
The calculations were performed with the same parameters as in Fig.~\ref{fig:megaTcfigure}a.
}
\end{figure}

\end{appendix}
\end{widetext}

\bibliography{BLGSC.bib}

\begin{thebibliography}{51}%
\makeatletter
\providecommand \@ifxundefined [1]{%
 \@ifx{#1\undefined}
}%
\providecommand \@ifnum [1]{%
 \ifnum #1\expandafter \@firstoftwo
 \else \expandafter \@secondoftwo
 \fi
}%
\providecommand \@ifx [1]{%
 \ifx #1\expandafter \@firstoftwo
 \else \expandafter \@secondoftwo
 \fi
}%
\providecommand \natexlab [1]{#1}%
\providecommand \enquote  [1]{``#1''}%
\providecommand \bibnamefont  [1]{#1}%
\providecommand \bibfnamefont [1]{#1}%
\providecommand \citenamefont [1]{#1}%
\providecommand \href@noop [0]{\@secondoftwo}%
\providecommand \href [0]{\begingroup \@sanitize@url \@href}%
\providecommand \@href[1]{\@@startlink{#1}\@@href}%
\providecommand \@@href[1]{\endgroup#1\@@endlink}%
\providecommand \@sanitize@url [0]{\catcode `\\12\catcode `\$12\catcode
  `\&12\catcode `\#12\catcode `\^12\catcode `\_12\catcode `\%12\relax}%
\providecommand \@@startlink[1]{}%
\providecommand \@@endlink[0]{}%
\providecommand \url  [0]{\begingroup\@sanitize@url \@url }%
\providecommand \@url [1]{\endgroup\@href {#1}{\urlprefix }}%
\providecommand \urlprefix  [0]{URL }%
\providecommand \Eprint [0]{\href }%
\providecommand \doibase [0]{https://doi.org/}%
\providecommand \selectlanguage [0]{\@gobble}%
\providecommand \bibinfo  [0]{\@secondoftwo}%
\providecommand \bibfield  [0]{\@secondoftwo}%
\providecommand \translation [1]{[#1]}%
\providecommand \BibitemOpen [0]{}%
\providecommand \bibitemStop [0]{}%
\providecommand \bibitemNoStop [0]{.\EOS\space}%
\providecommand \EOS [0]{\spacefactor3000\relax}%
\providecommand \BibitemShut  [1]{\csname bibitem#1\endcsname}%
\let\auto@bib@innerbib\@empty
\bibitem [{\citenamefont {Ginzburg}\ and\ \citenamefont
  {Landau}(1950)}]{GinzburgLandau}%
  \BibitemOpen
  \bibfield  {author} {\bibinfo {author} {\bibfnamefont {V.~L.}\ \bibnamefont
  {Ginzburg}}\ and\ \bibinfo {author} {\bibfnamefont {L.~D.}\ \bibnamefont
  {Landau}},\ }\bibfield  {title} {\bibinfo {title} {{On the Theory of
  superconductivity}},\ }\href
  {https://doi.org/10.1016/B978-0-08-010586-4.50035-3} {\bibfield  {journal}
  {\bibinfo  {journal} {Zh. Eksp. Teor. Fiz.}\ }\textbf {\bibinfo {volume}
  {20}},\ \bibinfo {pages} {1064} (\bibinfo {year} {1950})}\BibitemShut
  {NoStop}%
\bibitem [{\citenamefont {Tinkham}(2004)}]{tinkham2004introduction}%
  \BibitemOpen
  \bibfield  {author} {\bibinfo {author} {\bibfnamefont {M.}~\bibnamefont
  {Tinkham}},\ }\href@noop {} {\emph {\bibinfo {title} {Introduction to
  superconductivity}}}\ (\bibinfo  {publisher} {Courier Corporation},\ \bibinfo
  {year} {2004})\BibitemShut {NoStop}%
\bibitem [{\citenamefont {Clogston}(1962{\natexlab{a}})}]{Clogston}%
  \BibitemOpen
  \bibfield  {author} {\bibinfo {author} {\bibfnamefont {A.~M.}\ \bibnamefont
  {Clogston}},\ }\bibfield  {title} {\bibinfo {title} {Upper limit for the
  critical field in hard superconductors},\ }\href
  {https://doi.org/10.1103/PhysRevLett.9.266} {\bibfield  {journal} {\bibinfo
  {journal} {Phys. Rev. Lett.}\ }\textbf {\bibinfo {volume} {9}},\ \bibinfo
  {pages} {266} (\bibinfo {year} {1962}{\natexlab{a}})}\BibitemShut {NoStop}%
\bibitem [{\citenamefont {Clogston}(1962{\natexlab{b}})}]{ClogsonLimit}%
  \BibitemOpen
  \bibfield  {author} {\bibinfo {author} {\bibfnamefont {A.~M.}\ \bibnamefont
  {Clogston}},\ }\bibfield  {title} {\bibinfo {title} {Upper limit for the
  critical field in hard superconductors},\ }\href
  {https://doi.org/10.1103/PhysRevLett.9.266} {\bibfield  {journal} {\bibinfo
  {journal} {Phys. Rev. Lett.}\ }\textbf {\bibinfo {volume} {9}},\ \bibinfo
  {pages} {266} (\bibinfo {year} {1962}{\natexlab{b}})}\BibitemShut {NoStop}%
\bibitem [{\citenamefont {Lu}\ \emph {et~al.}(2015)\citenamefont {Lu},
  \citenamefont {Zheliuk}, \citenamefont {Leermakers}, \citenamefont {Yuan},
  \citenamefont {Zeitler}, \citenamefont {Law},\ and\ \citenamefont
  {Ye}}]{IsingSuperconductivityMoS2}%
  \BibitemOpen
  \bibfield  {author} {\bibinfo {author} {\bibfnamefont {J.~M.}\ \bibnamefont
  {Lu}}, \bibinfo {author} {\bibfnamefont {O.}~\bibnamefont {Zheliuk}},
  \bibinfo {author} {\bibfnamefont {I.}~\bibnamefont {Leermakers}}, \bibinfo
  {author} {\bibfnamefont {N.~F.~Q.}\ \bibnamefont {Yuan}}, \bibinfo {author}
  {\bibfnamefont {U.}~\bibnamefont {Zeitler}}, \bibinfo {author} {\bibfnamefont
  {K.~T.}\ \bibnamefont {Law}},\ and\ \bibinfo {author} {\bibfnamefont {J.~T.}\
  \bibnamefont {Ye}},\ }\bibfield  {title} {\bibinfo {title} {Evidence for
  two-dimensional ising superconductivity in gated mos2},\ }\href
  {https://doi.org/10.1126/science.aab2277} {\bibfield  {journal} {\bibinfo
  {journal} {Science}\ }\textbf {\bibinfo {volume} {350}},\ \bibinfo {pages}
  {1353} (\bibinfo {year} {2015})},\ \Eprint
  {https://arxiv.org/abs/https://science.sciencemag.org/content/350/6266/1353.full.pdf}
  {https://science.sciencemag.org/content/350/6266/1353.full.pdf} \BibitemShut
  {NoStop}%
\bibitem [{\citenamefont {Xi}\ \emph {et~al.}(2016)\citenamefont {Xi},
  \citenamefont {Wang}, \citenamefont {Zhao}, \citenamefont {Park},
  \citenamefont {Law}, \citenamefont {Berger}, \citenamefont {Forr{\'o}},
  \citenamefont {Shan},\ and\ \citenamefont
  {Mak}}]{IsingSuperconductivityNbSe2}%
  \BibitemOpen
  \bibfield  {author} {\bibinfo {author} {\bibfnamefont {X.}~\bibnamefont
  {Xi}}, \bibinfo {author} {\bibfnamefont {Z.}~\bibnamefont {Wang}}, \bibinfo
  {author} {\bibfnamefont {W.}~\bibnamefont {Zhao}}, \bibinfo {author}
  {\bibfnamefont {J.-H.}\ \bibnamefont {Park}}, \bibinfo {author}
  {\bibfnamefont {K.~T.}\ \bibnamefont {Law}}, \bibinfo {author} {\bibfnamefont
  {H.}~\bibnamefont {Berger}}, \bibinfo {author} {\bibfnamefont
  {L.}~\bibnamefont {Forr{\'o}}}, \bibinfo {author} {\bibfnamefont
  {J.}~\bibnamefont {Shan}},\ and\ \bibinfo {author} {\bibfnamefont {K.~F.}\
  \bibnamefont {Mak}},\ }\bibfield  {title} {\bibinfo {title} {Ising pairing in
  superconducting nbse2 atomic layers},\ }\href
  {https://doi.org/10.1038/nphys3538} {\bibfield  {journal} {\bibinfo
  {journal} {Nature Physics}\ }\textbf {\bibinfo {volume} {12}},\ \bibinfo
  {pages} {139} (\bibinfo {year} {2016})}\BibitemShut {NoStop}%
\bibitem [{\citenamefont {de~la Barrera}\ \emph {et~al.}(2018)\citenamefont
  {de~la Barrera}, \citenamefont {Sinko}, \citenamefont {Gopalan},
  \citenamefont {Sivadas}, \citenamefont {Seyler}, \citenamefont {Watanabe},
  \citenamefont {Taniguchi}, \citenamefont {Tsen}, \citenamefont {Xu},
  \citenamefont {Xiao},\ and\ \citenamefont
  {Hunt}}]{delaBarrera2018IsingSuperconductivityTas2Nbse2TUNING}%
  \BibitemOpen
  \bibfield  {author} {\bibinfo {author} {\bibfnamefont {S.~C.}\ \bibnamefont
  {de~la Barrera}}, \bibinfo {author} {\bibfnamefont {M.~R.}\ \bibnamefont
  {Sinko}}, \bibinfo {author} {\bibfnamefont {D.~P.}\ \bibnamefont {Gopalan}},
  \bibinfo {author} {\bibfnamefont {N.}~\bibnamefont {Sivadas}}, \bibinfo
  {author} {\bibfnamefont {K.~L.}\ \bibnamefont {Seyler}}, \bibinfo {author}
  {\bibfnamefont {K.}~\bibnamefont {Watanabe}}, \bibinfo {author}
  {\bibfnamefont {T.}~\bibnamefont {Taniguchi}}, \bibinfo {author}
  {\bibfnamefont {A.~W.}\ \bibnamefont {Tsen}}, \bibinfo {author}
  {\bibfnamefont {X.}~\bibnamefont {Xu}}, \bibinfo {author} {\bibfnamefont
  {D.}~\bibnamefont {Xiao}},\ and\ \bibinfo {author} {\bibfnamefont {B.~M.}\
  \bibnamefont {Hunt}},\ }\bibfield  {title} {\bibinfo {title} {Tuning ising
  superconductivity with layer and spin--orbit coupling in two-dimensional
  transition-metal dichalcogenides},\ }\href
  {https://doi.org/10.1038/s41467-018-03888-4} {\bibfield  {journal} {\bibinfo
  {journal} {Nature Communications}\ }\textbf {\bibinfo {volume} {9}},\
  \bibinfo {pages} {1427} (\bibinfo {year} {2018})}\BibitemShut {NoStop}%
\bibitem [{\citenamefont {Cao}\ \emph {et~al.}(2021)\citenamefont {Cao},
  \citenamefont {Park}, \citenamefont {Watanabe}, \citenamefont {Taniguchi},\
  and\ \citenamefont {Jarillo-Herrero}}]{Cao2021PauliLimitViolationTrilayer}%
  \BibitemOpen
  \bibfield  {author} {\bibinfo {author} {\bibfnamefont {Y.}~\bibnamefont
  {Cao}}, \bibinfo {author} {\bibfnamefont {J.~M.}\ \bibnamefont {Park}},
  \bibinfo {author} {\bibfnamefont {K.}~\bibnamefont {Watanabe}}, \bibinfo
  {author} {\bibfnamefont {T.}~\bibnamefont {Taniguchi}},\ and\ \bibinfo
  {author} {\bibfnamefont {P.}~\bibnamefont {Jarillo-Herrero}},\ }\bibfield
  {title} {\bibinfo {title} {Pauli-limit violation and re-entrant
  superconductivity in moir{\'e} graphene},\ }\href
  {https://doi.org/10.1038/s41586-021-03685-y} {\bibfield  {journal} {\bibinfo
  {journal} {Nature}\ }\textbf {\bibinfo {volume} {595}},\ \bibinfo {pages}
  {526} (\bibinfo {year} {2021})}\BibitemShut {NoStop}%
\bibitem [{\citenamefont {Park}\ \emph {et~al.}(2022)\citenamefont {Park},
  \citenamefont {Cao}, \citenamefont {Xia}, \citenamefont {Sun}, \citenamefont
  {Watanabe}, \citenamefont {Taniguchi},\ and\ \citenamefont
  {Jarillo-Herrero}}]{Park2022MATngFamily}%
  \BibitemOpen
  \bibfield  {author} {\bibinfo {author} {\bibfnamefont {J.~M.}\ \bibnamefont
  {Park}}, \bibinfo {author} {\bibfnamefont {Y.}~\bibnamefont {Cao}}, \bibinfo
  {author} {\bibfnamefont {L.-Q.}\ \bibnamefont {Xia}}, \bibinfo {author}
  {\bibfnamefont {S.}~\bibnamefont {Sun}}, \bibinfo {author} {\bibfnamefont
  {K.}~\bibnamefont {Watanabe}}, \bibinfo {author} {\bibfnamefont
  {T.}~\bibnamefont {Taniguchi}},\ and\ \bibinfo {author} {\bibfnamefont
  {P.}~\bibnamefont {Jarillo-Herrero}},\ }\bibfield  {title} {\bibinfo {title}
  {Robust superconductivity in magic-angle multilayer graphene family},\ }\href
  {https://doi.org/10.1038/s41563-022-01287-1} {\bibfield  {journal} {\bibinfo
  {journal} {Nature Materials}\ }\textbf {\bibinfo {volume} {21}},\ \bibinfo
  {pages} {877} (\bibinfo {year} {2022})}\BibitemShut {NoStop}%
\bibitem [{\citenamefont {Zhang}\ \emph
  {et~al.}(2022{\natexlab{a}})\citenamefont {Zhang}, \citenamefont {Polski},
  \citenamefont {Lewandowski}, \citenamefont {Thomson}, \citenamefont {Peng},
  \citenamefont {Choi}, \citenamefont {Kim}, \citenamefont {Watanabe},
  \citenamefont {Taniguchi}, \citenamefont {Alicea}, \citenamefont {von Oppen},
  \citenamefont {Refael},\ and\ \citenamefont
  {Nadj-Perge}}]{NadjPergAscendenceMATng}%
  \BibitemOpen
  \bibfield  {author} {\bibinfo {author} {\bibfnamefont {Y.}~\bibnamefont
  {Zhang}}, \bibinfo {author} {\bibfnamefont {R.}~\bibnamefont {Polski}},
  \bibinfo {author} {\bibfnamefont {C.}~\bibnamefont {Lewandowski}}, \bibinfo
  {author} {\bibfnamefont {A.}~\bibnamefont {Thomson}}, \bibinfo {author}
  {\bibfnamefont {Y.}~\bibnamefont {Peng}}, \bibinfo {author} {\bibfnamefont
  {Y.}~\bibnamefont {Choi}}, \bibinfo {author} {\bibfnamefont {H.}~\bibnamefont
  {Kim}}, \bibinfo {author} {\bibfnamefont {K.}~\bibnamefont {Watanabe}},
  \bibinfo {author} {\bibfnamefont {T.}~\bibnamefont {Taniguchi}}, \bibinfo
  {author} {\bibfnamefont {J.}~\bibnamefont {Alicea}}, \bibinfo {author}
  {\bibfnamefont {F.}~\bibnamefont {von Oppen}}, \bibinfo {author}
  {\bibfnamefont {G.}~\bibnamefont {Refael}},\ and\ \bibinfo {author}
  {\bibfnamefont {S.}~\bibnamefont {Nadj-Perge}},\ }\bibfield  {title}
  {\bibinfo {title} {Promotion of superconductivity in magic-angle graphene
  multilayers},\ }\href {https://doi.org/10.1126/science.abn8585} {\bibfield
  {journal} {\bibinfo  {journal} {Science}\ }\textbf {\bibinfo {volume}
  {377}},\ \bibinfo {pages} {1538} (\bibinfo {year} {2022}{\natexlab{a}})},\
  \Eprint
  {https://arxiv.org/abs/https://www.science.org/doi/pdf/10.1126/science.abn8585}
  {https://www.science.org/doi/pdf/10.1126/science.abn8585} \BibitemShut
  {NoStop}%
\bibitem [{\citenamefont {Zhou}\ \emph {et~al.}(2022)\citenamefont {Zhou},
  \citenamefont {Holleis}, \citenamefont {Saito}, \citenamefont {Cohen},
  \citenamefont {Huynh}, \citenamefont {Patterson}, \citenamefont {Yang},
  \citenamefont {Taniguchi}, \citenamefont {Watanabe},\ and\ \citenamefont
  {Young}}]{ZhouYoungBLGZeeman}%
  \BibitemOpen
  \bibfield  {author} {\bibinfo {author} {\bibfnamefont {H.}~\bibnamefont
  {Zhou}}, \bibinfo {author} {\bibfnamefont {L.}~\bibnamefont {Holleis}},
  \bibinfo {author} {\bibfnamefont {Y.}~\bibnamefont {Saito}}, \bibinfo
  {author} {\bibfnamefont {L.}~\bibnamefont {Cohen}}, \bibinfo {author}
  {\bibfnamefont {W.}~\bibnamefont {Huynh}}, \bibinfo {author} {\bibfnamefont
  {C.~L.}\ \bibnamefont {Patterson}}, \bibinfo {author} {\bibfnamefont
  {F.}~\bibnamefont {Yang}}, \bibinfo {author} {\bibfnamefont {T.}~\bibnamefont
  {Taniguchi}}, \bibinfo {author} {\bibfnamefont {K.}~\bibnamefont
  {Watanabe}},\ and\ \bibinfo {author} {\bibfnamefont {A.~F.}\ \bibnamefont
  {Young}},\ }\bibfield  {title} {\bibinfo {title} {Isospin magnetism and
  spin-polarized superconductivity in bernal bilayer graphene},\ }\href
  {https://doi.org/10.1126/science.abm8386} {\bibfield  {journal} {\bibinfo
  {journal} {Science}\ }\textbf {\bibinfo {volume} {375}},\ \bibinfo {pages}
  {774} (\bibinfo {year} {2022})},\ \Eprint
  {https://arxiv.org/abs/https://www.science.org/doi/pdf/10.1126/science.abm8386}
  {https://www.science.org/doi/pdf/10.1126/science.abm8386} \BibitemShut
  {NoStop}%
\bibitem [{\citenamefont {Saito}\ \emph {et~al.}(2020)\citenamefont {Saito},
  \citenamefont {Ge}, \citenamefont {Watanabe}, \citenamefont {Taniguchi},\
  and\ \citenamefont {Young}}]{YoungTuningSC}%
  \BibitemOpen
  \bibfield  {author} {\bibinfo {author} {\bibfnamefont {Y.}~\bibnamefont
  {Saito}}, \bibinfo {author} {\bibfnamefont {J.}~\bibnamefont {Ge}}, \bibinfo
  {author} {\bibfnamefont {K.}~\bibnamefont {Watanabe}}, \bibinfo {author}
  {\bibfnamefont {T.}~\bibnamefont {Taniguchi}},\ and\ \bibinfo {author}
  {\bibfnamefont {A.~F.}\ \bibnamefont {Young}},\ }\bibfield  {title} {\bibinfo
  {title} {Independent superconductors and correlated insulators in twisted
  bilayer graphene},\ }\href {https://doi.org/10.1038/s41567-020-0928-3}
  {\bibfield  {journal} {\bibinfo  {journal} {Nature Phys.}\ }\textbf {\bibinfo
  {volume} {16}},\ \bibinfo {pages} {926} (\bibinfo {year} {2020})}\BibitemShut
  {NoStop}%
\bibitem [{\citenamefont {Stepanov}\ \emph {et~al.}(2020)\citenamefont
  {Stepanov}, \citenamefont {Das}, \citenamefont {Lu}, \citenamefont
  {Fahimniya}, \citenamefont {Watanabe}, \citenamefont {Taniguchi},
  \citenamefont {Koppens}, \citenamefont {Lischner}, \citenamefont {Levitov},\
  and\ \citenamefont {Efetov}}]{EfetovTuningSC}%
  \BibitemOpen
  \bibfield  {author} {\bibinfo {author} {\bibfnamefont {P.}~\bibnamefont
  {Stepanov}}, \bibinfo {author} {\bibfnamefont {I.}~\bibnamefont {Das}},
  \bibinfo {author} {\bibfnamefont {X.}~\bibnamefont {Lu}}, \bibinfo {author}
  {\bibfnamefont {A.}~\bibnamefont {Fahimniya}}, \bibinfo {author}
  {\bibfnamefont {K.}~\bibnamefont {Watanabe}}, \bibinfo {author}
  {\bibfnamefont {T.}~\bibnamefont {Taniguchi}}, \bibinfo {author}
  {\bibfnamefont {F.~H.~L.}\ \bibnamefont {Koppens}}, \bibinfo {author}
  {\bibfnamefont {J.}~\bibnamefont {Lischner}}, \bibinfo {author}
  {\bibfnamefont {L.}~\bibnamefont {Levitov}},\ and\ \bibinfo {author}
  {\bibfnamefont {D.~K.}\ \bibnamefont {Efetov}},\ }\bibfield  {title}
  {\bibinfo {title} {Untying the insulating and superconducting orders in
  magic-angle graphene},\ }\href {https://doi.org/10.1038/s41586-020-2459-6}
  {\bibfield  {journal} {\bibinfo  {journal} {Nature}\ }\textbf {\bibinfo
  {volume} {583}},\ \bibinfo {pages} {375} (\bibinfo {year}
  {2020})}\BibitemShut {NoStop}%
\bibitem [{\citenamefont {Liu}\ \emph {et~al.}(2021)\citenamefont {Liu},
  \citenamefont {Wang}, \citenamefont {Watanabe}, \citenamefont {Taniguchi},
  \citenamefont {Vafek},\ and\ \citenamefont {Li}}]{BLGscreening}%
  \BibitemOpen
  \bibfield  {author} {\bibinfo {author} {\bibfnamefont {X.}~\bibnamefont
  {Liu}}, \bibinfo {author} {\bibfnamefont {Z.}~\bibnamefont {Wang}}, \bibinfo
  {author} {\bibfnamefont {K.}~\bibnamefont {Watanabe}}, \bibinfo {author}
  {\bibfnamefont {T.}~\bibnamefont {Taniguchi}}, \bibinfo {author}
  {\bibfnamefont {O.}~\bibnamefont {Vafek}},\ and\ \bibinfo {author}
  {\bibfnamefont {J.~I.~A.}\ \bibnamefont {Li}},\ }\bibfield  {title} {\bibinfo
  {title} {Tuning electron correlation in magic-angle twisted bilayer graphene
  using coulomb screening},\ }\href {https://doi.org/10.1126/science.abb8754}
  {\bibfield  {journal} {\bibinfo  {journal} {Science}\ }\textbf {\bibinfo
  {volume} {371}},\ \bibinfo {pages} {1261} (\bibinfo {year}
  {2021})}\BibitemShut {NoStop}%
\bibitem [{\citenamefont {Zhang}\ \emph
  {et~al.}(2022{\natexlab{b}})\citenamefont {Zhang}, \citenamefont {Polski},
  \citenamefont {Thomson}, \citenamefont {Lantagne-Hurtubise}, \citenamefont
  {Lewandowski}, \citenamefont {Zhou}, \citenamefont {Watanabe}, \citenamefont
  {Taniguchi}, \citenamefont {Alicea},\ and\ \citenamefont
  {Nadj-Perge}}]{ZhangBLGSOC}%
  \BibitemOpen
  \bibfield  {author} {\bibinfo {author} {\bibfnamefont {Y.}~\bibnamefont
  {Zhang}}, \bibinfo {author} {\bibfnamefont {R.}~\bibnamefont {Polski}},
  \bibinfo {author} {\bibfnamefont {A.}~\bibnamefont {Thomson}}, \bibinfo
  {author} {\bibfnamefont {Ã.}~\bibnamefont {Lantagne-Hurtubise}}, \bibinfo
  {author} {\bibfnamefont {C.}~\bibnamefont {Lewandowski}}, \bibinfo {author}
  {\bibfnamefont {H.}~\bibnamefont {Zhou}}, \bibinfo {author} {\bibfnamefont
  {K.}~\bibnamefont {Watanabe}}, \bibinfo {author} {\bibfnamefont
  {T.}~\bibnamefont {Taniguchi}}, \bibinfo {author} {\bibfnamefont
  {J.}~\bibnamefont {Alicea}},\ and\ \bibinfo {author} {\bibfnamefont
  {S.}~\bibnamefont {Nadj-Perge}},\ }\href
  {https://doi.org/10.48550/ARXIV.2205.05087} {\bibinfo {title} {Spin-orbit
  enhanced superconductivity in bernal bilayer graphene}} (\bibinfo {year}
  {2022}{\natexlab{b}})\BibitemShut {NoStop}%
\bibitem [{\citenamefont {Holleis}\ \emph
  {et~al.}(2023{\natexlab{a}})\citenamefont {Holleis}, \citenamefont
  {Patterson}, \citenamefont {Zhang}, \citenamefont {Yoo}, \citenamefont
  {Zhou}, \citenamefont {Taniguchi}, \citenamefont {Watanabe}, \citenamefont
  {Nadj-Perge},\ and\ \citenamefont {Young}}]{BLGyoungNadjperge}%
  \BibitemOpen
  \bibfield  {author} {\bibinfo {author} {\bibfnamefont {L.}~\bibnamefont
  {Holleis}}, \bibinfo {author} {\bibfnamefont {C.~L.}\ \bibnamefont
  {Patterson}}, \bibinfo {author} {\bibfnamefont {Y.}~\bibnamefont {Zhang}},
  \bibinfo {author} {\bibfnamefont {H.~M.}\ \bibnamefont {Yoo}}, \bibinfo
  {author} {\bibfnamefont {H.}~\bibnamefont {Zhou}}, \bibinfo {author}
  {\bibfnamefont {T.}~\bibnamefont {Taniguchi}}, \bibinfo {author}
  {\bibfnamefont {K.}~\bibnamefont {Watanabe}}, \bibinfo {author}
  {\bibfnamefont {S.}~\bibnamefont {Nadj-Perge}},\ and\ \bibinfo {author}
  {\bibfnamefont {A.~F.}\ \bibnamefont {Young}},\ }\href
  {https://doi.org/10.48550/ARXIV.2303.00742} {\bibinfo {title} {Ising
  superconductivity and nematicity in bernal bilayer graphene with strong spin
  orbit coupling}} (\bibinfo {year} {2023}{\natexlab{a}})\BibitemShut {NoStop}%
\bibitem [{\citenamefont {Anderson}(1959)}]{ANDERSONtheorem}%
  \BibitemOpen
  \bibfield  {author} {\bibinfo {author} {\bibfnamefont {P.}~\bibnamefont
  {Anderson}},\ }\bibfield  {title} {\bibinfo {title} {Theory of dirty
  superconductors},\ }\href
  {https://doi.org/https://doi.org/10.1016/0022-3697(59)90036-8} {\bibfield
  {journal} {\bibinfo  {journal} {Journal of Physics and Chemistry of Solids}\
  }\textbf {\bibinfo {volume} {11}},\ \bibinfo {pages} {26} (\bibinfo {year}
  {1959})}\BibitemShut {NoStop}%
\bibitem [{\citenamefont {McCann}\ and\ \citenamefont
  {Fal'ko}(2006)}]{MccanBernalGrapheneLandauLevel}%
  \BibitemOpen
  \bibfield  {author} {\bibinfo {author} {\bibfnamefont {E.}~\bibnamefont
  {McCann}}\ and\ \bibinfo {author} {\bibfnamefont {V.~I.}\ \bibnamefont
  {Fal'ko}},\ }\bibfield  {title} {\bibinfo {title} {Landau-level degeneracy
  and quantum hall effect in a graphite bilayer},\ }\href
  {https://doi.org/10.1103/PhysRevLett.96.086805} {\bibfield  {journal}
  {\bibinfo  {journal} {Phys. Rev. Lett.}\ }\textbf {\bibinfo {volume} {96}},\
  \bibinfo {pages} {086805} (\bibinfo {year} {2006})}\BibitemShut {NoStop}%
\bibitem [{\citenamefont {McCann}\ and\ \citenamefont
  {Koshino}(2013)}]{McCannBilayerGrapheneBernal}%
  \BibitemOpen
  \bibfield  {author} {\bibinfo {author} {\bibfnamefont {E.}~\bibnamefont
  {McCann}}\ and\ \bibinfo {author} {\bibfnamefont {M.}~\bibnamefont
  {Koshino}},\ }\bibfield  {title} {\bibinfo {title} {The electronic properties
  of bilayer graphene},\ }\href {https://doi.org/10.1088/0034-4885/76/5/056503}
  {\bibfield  {journal} {\bibinfo  {journal} {Reports on Progress in Physics}\
  }\textbf {\bibinfo {volume} {76}},\ \bibinfo {pages} {056503} (\bibinfo
  {year} {2013})}\BibitemShut {NoStop}%
\bibitem [{\citenamefont {Zhou}\ \emph {et~al.}(2021)\citenamefont {Zhou},
  \citenamefont {Xie}, \citenamefont {Ghazaryan}, \citenamefont {Holder},
  \citenamefont {Ehrets}, \citenamefont {Spanton}, \citenamefont {Taniguchi},
  \citenamefont {Watanabe}, \citenamefont {Berg}, \citenamefont {Serbyn},\ and\
  \citenamefont {Young}}]{Zhou2021RTGYoung}%
  \BibitemOpen
  \bibfield  {author} {\bibinfo {author} {\bibfnamefont {H.}~\bibnamefont
  {Zhou}}, \bibinfo {author} {\bibfnamefont {T.}~\bibnamefont {Xie}}, \bibinfo
  {author} {\bibfnamefont {A.}~\bibnamefont {Ghazaryan}}, \bibinfo {author}
  {\bibfnamefont {T.}~\bibnamefont {Holder}}, \bibinfo {author} {\bibfnamefont
  {J.~R.}\ \bibnamefont {Ehrets}}, \bibinfo {author} {\bibfnamefont {E.~M.}\
  \bibnamefont {Spanton}}, \bibinfo {author} {\bibfnamefont {T.}~\bibnamefont
  {Taniguchi}}, \bibinfo {author} {\bibfnamefont {K.}~\bibnamefont {Watanabe}},
  \bibinfo {author} {\bibfnamefont {E.}~\bibnamefont {Berg}}, \bibinfo {author}
  {\bibfnamefont {M.}~\bibnamefont {Serbyn}},\ and\ \bibinfo {author}
  {\bibfnamefont {A.~F.}\ \bibnamefont {Young}},\ }\bibfield  {title} {\bibinfo
  {title} {Half- and quarter-metals in rhombohedral trilayer graphene},\ }\href
  {https://doi.org/10.1038/s41586-021-03938-w} {\bibfield  {journal} {\bibinfo
  {journal} {Nature}\ }\textbf {\bibinfo {volume} {598}},\ \bibinfo {pages}
  {429} (\bibinfo {year} {2021})}\BibitemShut {NoStop}%
\bibitem [{Note1()}]{Note1}%
  \BibitemOpen
  \bibinfo {note} {The shifting phase boundary as a function of magnetic field
  and density seen in Ref.~\cite {ZhouYoungBLGZeeman} is suggestive of linear
  coupling between the magnetic field and the order parameter (see Sec.~\ref
  {sec:softeningtransitionsmagnetization} and Fig.~\ref
  {fig:softenedtransition}a). As such, the possibility of spontaneous
  spin-polarization and Zeeman coupling of the electrons to the in-plane
  magnetic field is naturally the most appealing.}\BibitemShut {Stop}%
\bibitem [{\citenamefont {Zondiner}\ \emph {et~al.}(2020)\citenamefont
  {Zondiner}, \citenamefont {Rozen}, \citenamefont {Rodan-Legrain},
  \citenamefont {Cao}, \citenamefont {Queiroz}, \citenamefont {Taniguchi},
  \citenamefont {Watanabe}, \citenamefont {Oreg}, \citenamefont {von Oppen},
  \citenamefont {Stern}, \citenamefont {Berg}, \citenamefont
  {Jarillo-Herrero},\ and\ \citenamefont {Ilani}}]{DiracRevivals}%
  \BibitemOpen
  \bibfield  {author} {\bibinfo {author} {\bibfnamefont {U.}~\bibnamefont
  {Zondiner}}, \bibinfo {author} {\bibfnamefont {A.}~\bibnamefont {Rozen}},
  \bibinfo {author} {\bibfnamefont {D.}~\bibnamefont {Rodan-Legrain}}, \bibinfo
  {author} {\bibfnamefont {Y.}~\bibnamefont {Cao}}, \bibinfo {author}
  {\bibfnamefont {R.}~\bibnamefont {Queiroz}}, \bibinfo {author} {\bibfnamefont
  {T.}~\bibnamefont {Taniguchi}}, \bibinfo {author} {\bibfnamefont
  {K.}~\bibnamefont {Watanabe}}, \bibinfo {author} {\bibfnamefont
  {Y.}~\bibnamefont {Oreg}}, \bibinfo {author} {\bibfnamefont {F.}~\bibnamefont
  {von Oppen}}, \bibinfo {author} {\bibfnamefont {A.}~\bibnamefont {Stern}},
  \bibinfo {author} {\bibfnamefont {E.}~\bibnamefont {Berg}}, \bibinfo {author}
  {\bibfnamefont {P.}~\bibnamefont {Jarillo-Herrero}},\ and\ \bibinfo {author}
  {\bibfnamefont {S.}~\bibnamefont {Ilani}},\ }\bibfield  {title} {\bibinfo
  {title} {Cascade of phase transitions and dirac revivals in magic-angle
  graphene},\ }\href {https://doi.org/10.1038/s41586-020-2373-y} {\bibfield
  {journal} {\bibinfo  {journal} {Nature}\ }\textbf {\bibinfo {volume} {582}},\
  \bibinfo {pages} {203} (\bibinfo {year} {2020})}\BibitemShut {NoStop}%
\bibitem [{\citenamefont {Shavit}\ \emph {et~al.}(2021)\citenamefont {Shavit},
  \citenamefont {Berg}, \citenamefont {Stern},\ and\ \citenamefont
  {Oreg}}]{ShavitMaATBGprl}%
  \BibitemOpen
  \bibfield  {author} {\bibinfo {author} {\bibfnamefont {G.}~\bibnamefont
  {Shavit}}, \bibinfo {author} {\bibfnamefont {E.}~\bibnamefont {Berg}},
  \bibinfo {author} {\bibfnamefont {A.}~\bibnamefont {Stern}},\ and\ \bibinfo
  {author} {\bibfnamefont {Y.}~\bibnamefont {Oreg}},\ }\bibfield  {title}
  {\bibinfo {title} {Theory of correlated insulators and superconductivity in
  twisted bilayer graphene},\ }\href
  {https://doi.org/10.1103/PhysRevLett.127.247703} {\bibfield  {journal}
  {\bibinfo  {journal} {Phys. Rev. Lett.}\ }\textbf {\bibinfo {volume} {127}},\
  \bibinfo {pages} {247703} (\bibinfo {year} {2021})}\BibitemShut {NoStop}%
\bibitem [{\citenamefont {Holleis}\ \emph
  {et~al.}(2023{\natexlab{b}})\citenamefont {Holleis}, \citenamefont
  {Patterson}, \citenamefont {Zhang}, \citenamefont {Yoo}, \citenamefont
  {Zhou}, \citenamefont {Taniguchi}, \citenamefont {Watanabe}, \citenamefont
  {Nadj-Perge},\ and\ \citenamefont {Young}}]{YoungPerrgeBLGcompressibility}%
  \BibitemOpen
  \bibfield  {author} {\bibinfo {author} {\bibfnamefont {L.}~\bibnamefont
  {Holleis}}, \bibinfo {author} {\bibfnamefont {C.~L.}\ \bibnamefont
  {Patterson}}, \bibinfo {author} {\bibfnamefont {Y.}~\bibnamefont {Zhang}},
  \bibinfo {author} {\bibfnamefont {H.~M.}\ \bibnamefont {Yoo}}, \bibinfo
  {author} {\bibfnamefont {H.}~\bibnamefont {Zhou}}, \bibinfo {author}
  {\bibfnamefont {T.}~\bibnamefont {Taniguchi}}, \bibinfo {author}
  {\bibfnamefont {K.}~\bibnamefont {Watanabe}}, \bibinfo {author}
  {\bibfnamefont {S.}~\bibnamefont {Nadj-Perge}},\ and\ \bibinfo {author}
  {\bibfnamefont {A.~F.}\ \bibnamefont {Young}},\ }\href@noop {} {\bibinfo
  {title} {Ising superconductivity and nematicity in bernal bilayer graphene
  with strong spin orbit coupling}} (\bibinfo {year} {2023}{\natexlab{b}}),\
  \Eprint {https://arxiv.org/abs/2303.00742} {arXiv:2303.00742
  [cond-mat.supr-con]} \BibitemShut {NoStop}%
\bibitem [{\citenamefont {Chou}\ \emph
  {et~al.}(2022{\natexlab{a}})\citenamefont {Chou}, \citenamefont {Wu},
  \citenamefont {Sau},\ and\ \citenamefont {Das~Sarma}}]{DasSarmaBLGphonons}%
  \BibitemOpen
  \bibfield  {author} {\bibinfo {author} {\bibfnamefont {Y.-Z.}\ \bibnamefont
  {Chou}}, \bibinfo {author} {\bibfnamefont {F.}~\bibnamefont {Wu}}, \bibinfo
  {author} {\bibfnamefont {J.~D.}\ \bibnamefont {Sau}},\ and\ \bibinfo {author}
  {\bibfnamefont {S.}~\bibnamefont {Das~Sarma}},\ }\bibfield  {title} {\bibinfo
  {title} {Acoustic-phonon-mediated superconductivity in bernal bilayer
  graphene},\ }\href {https://doi.org/10.1103/PhysRevB.105.L100503} {\bibfield
  {journal} {\bibinfo  {journal} {Phys. Rev. B}\ }\textbf {\bibinfo {volume}
  {105}},\ \bibinfo {pages} {L100503} (\bibinfo {year}
  {2022}{\natexlab{a}})}\BibitemShut {NoStop}%
\bibitem [{\citenamefont {Lian}\ \emph {et~al.}(2019)\citenamefont {Lian},
  \citenamefont {Wang},\ and\ \citenamefont
  {Bernevig}}]{LianBioBernevigPRLphonons}%
  \BibitemOpen
  \bibfield  {author} {\bibinfo {author} {\bibfnamefont {B.}~\bibnamefont
  {Lian}}, \bibinfo {author} {\bibfnamefont {Z.}~\bibnamefont {Wang}},\ and\
  \bibinfo {author} {\bibfnamefont {B.~A.}\ \bibnamefont {Bernevig}},\
  }\bibfield  {title} {\bibinfo {title} {Twisted bilayer graphene: A
  phonon-driven superconductor},\ }\href
  {https://doi.org/10.1103/PhysRevLett.122.257002} {\bibfield  {journal}
  {\bibinfo  {journal} {Phys. Rev. Lett.}\ }\textbf {\bibinfo {volume} {122}},\
  \bibinfo {pages} {257002} (\bibinfo {year} {2019})}\BibitemShut {NoStop}%
\bibitem [{\citenamefont {Kheirabadi}\ \emph {et~al.}(2016)\citenamefont
  {Kheirabadi}, \citenamefont {McCann},\ and\ \citenamefont
  {Fal'ko}}]{BLGmagneticfieldorbital}%
  \BibitemOpen
  \bibfield  {author} {\bibinfo {author} {\bibfnamefont {N.}~\bibnamefont
  {Kheirabadi}}, \bibinfo {author} {\bibfnamefont {E.}~\bibnamefont {McCann}},\
  and\ \bibinfo {author} {\bibfnamefont {V.~I.}\ \bibnamefont {Fal'ko}},\
  }\bibfield  {title} {\bibinfo {title} {Magnetic ratchet effect in bilayer
  graphene},\ }\href {https://doi.org/10.1103/PhysRevB.94.165404} {\bibfield
  {journal} {\bibinfo  {journal} {Phys. Rev. B}\ }\textbf {\bibinfo {volume}
  {94}},\ \bibinfo {pages} {165404} (\bibinfo {year} {2016})}\BibitemShut
  {NoStop}%
\bibitem [{\citenamefont {Kol\'a\ifmmode~\check{r}\else \v{r}\fi{}}\ \emph
  {et~al.}(2023)\citenamefont {Kol\'a\ifmmode~\check{r}\else \v{r}\fi{}},
  \citenamefont {Shavit}, \citenamefont {Mora}, \citenamefont {Oreg},\ and\
  \citenamefont {von Oppen}}]{matbgKIVCanderson}%
  \BibitemOpen
  \bibfield  {author} {\bibinfo {author} {\bibfnamefont {K.~c.~v.}\
  \bibnamefont {Kol\'a\ifmmode~\check{r}\else \v{r}\fi{}}}, \bibinfo {author}
  {\bibfnamefont {G.}~\bibnamefont {Shavit}}, \bibinfo {author} {\bibfnamefont
  {C.}~\bibnamefont {Mora}}, \bibinfo {author} {\bibfnamefont {Y.}~\bibnamefont
  {Oreg}},\ and\ \bibinfo {author} {\bibfnamefont {F.}~\bibnamefont {von
  Oppen}},\ }\bibfield  {title} {\bibinfo {title} {Anderson's theorem for
  correlated insulating states in twisted bilayer graphene},\ }\href
  {https://doi.org/10.1103/PhysRevLett.130.076204} {\bibfield  {journal}
  {\bibinfo  {journal} {Phys. Rev. Lett.}\ }\textbf {\bibinfo {volume} {130}},\
  \bibinfo {pages} {076204} (\bibinfo {year} {2023})}\BibitemShut {NoStop}%
\bibitem [{\citenamefont {Abrikosov}\ and\ \citenamefont
  {Gorkov}(1959)}]{abrikosovandersonthorem}%
  \BibitemOpen
  \bibfield  {author} {\bibinfo {author} {\bibfnamefont {A.}~\bibnamefont
  {Abrikosov}}\ and\ \bibinfo {author} {\bibfnamefont {L.}~\bibnamefont
  {Gorkov}},\ }\bibfield  {title} {\bibinfo {title} {On the theory of
  superconducting alloys. 1. the electrodynamics of alloys at absolute zero},\
  }\href@noop {} {\bibfield  {journal} {\bibinfo  {journal} {Sov. Phys. JETP}\
  }\textbf {\bibinfo {volume} {8}},\ \bibinfo {pages} {1090} (\bibinfo {year}
  {1959})}\BibitemShut {NoStop}%
\bibitem [{\citenamefont {Maki}(1969)}]{maki1969gapless}%
  \BibitemOpen
  \bibfield  {author} {\bibinfo {author} {\bibfnamefont {K.}~\bibnamefont
  {Maki}},\ }\bibfield  {title} {\bibinfo {title} {Gapless superconductivity},\
  }in\ \href@noop {} {\emph {\bibinfo {booktitle} {Superconductivity}}},\
  Vol.~\bibinfo {volume} {2},\ \bibinfo {editor} {edited by\ \bibinfo {editor}
  {\bibfnamefont {R.~D.}\ \bibnamefont {Parks}}}\ (\bibinfo  {publisher}
  {Marcel Dekker},\ \bibinfo {year} {1969})\ p.\ \bibinfo {pages}
  {1035}\BibitemShut {NoStop}%
\bibitem [{\citenamefont {Ishigami}\ \emph {et~al.}(2007)\citenamefont
  {Ishigami}, \citenamefont {Chen}, \citenamefont {Cullen}, \citenamefont
  {Fuhrer},\ and\ \citenamefont {Williams}}]{ripplesIshigami}%
  \BibitemOpen
  \bibfield  {author} {\bibinfo {author} {\bibfnamefont {M.}~\bibnamefont
  {Ishigami}}, \bibinfo {author} {\bibfnamefont {J.~H.}\ \bibnamefont {Chen}},
  \bibinfo {author} {\bibfnamefont {W.~G.}\ \bibnamefont {Cullen}}, \bibinfo
  {author} {\bibfnamefont {M.~S.}\ \bibnamefont {Fuhrer}},\ and\ \bibinfo
  {author} {\bibfnamefont {E.~D.}\ \bibnamefont {Williams}},\ }\bibfield
  {title} {\bibinfo {title} {Atomic structure of graphene on sio2},\ }\href
  {https://doi.org/10.1021/nl070613a} {\bibfield  {journal} {\bibinfo
  {journal} {Nano Letters}\ }\textbf {\bibinfo {volume} {7}},\ \bibinfo {pages}
  {1643} (\bibinfo {year} {2007})},\ \bibinfo {note} {pMID: 17497819},\ \Eprint
  {https://arxiv.org/abs/https://doi.org/10.1021/nl070613a}
  {https://doi.org/10.1021/nl070613a} \BibitemShut {NoStop}%
\bibitem [{\citenamefont {Stolyarova}\ \emph {et~al.}(2007)\citenamefont
  {Stolyarova}, \citenamefont {Rim}, \citenamefont {Ryu}, \citenamefont
  {Maultzsch}, \citenamefont {Kim}, \citenamefont {Brus}, \citenamefont
  {Heinz}, \citenamefont {Hybertsen},\ and\ \citenamefont
  {Flynn}}]{ripplesStolyarova}%
  \BibitemOpen
  \bibfield  {author} {\bibinfo {author} {\bibfnamefont {E.}~\bibnamefont
  {Stolyarova}}, \bibinfo {author} {\bibfnamefont {K.~T.}\ \bibnamefont {Rim}},
  \bibinfo {author} {\bibfnamefont {S.}~\bibnamefont {Ryu}}, \bibinfo {author}
  {\bibfnamefont {J.}~\bibnamefont {Maultzsch}}, \bibinfo {author}
  {\bibfnamefont {P.}~\bibnamefont {Kim}}, \bibinfo {author} {\bibfnamefont
  {L.~E.}\ \bibnamefont {Brus}}, \bibinfo {author} {\bibfnamefont {T.~F.}\
  \bibnamefont {Heinz}}, \bibinfo {author} {\bibfnamefont {M.~S.}\ \bibnamefont
  {Hybertsen}},\ and\ \bibinfo {author} {\bibfnamefont {G.~W.}\ \bibnamefont
  {Flynn}},\ }\bibfield  {title} {\bibinfo {title} {High-resolution scanning
  tunneling microscopy imaging of mesoscopic graphene sheets on an insulating
  surface},\ }\href {https://doi.org/10.1073/pnas.0703337104} {\bibfield
  {journal} {\bibinfo  {journal} {Proceedings of the National Academy of
  Sciences}\ }\textbf {\bibinfo {volume} {104}},\ \bibinfo {pages} {9209}
  (\bibinfo {year} {2007})},\ \Eprint
  {https://arxiv.org/abs/https://www.pnas.org/doi/pdf/10.1073/pnas.0703337104}
  {https://www.pnas.org/doi/pdf/10.1073/pnas.0703337104} \BibitemShut {NoStop}%
\bibitem [{\citenamefont {Chen}\ \emph {et~al.}(2008)\citenamefont {Chen},
  \citenamefont {Jang}, \citenamefont {Adam}, \citenamefont {Fuhrer},
  \citenamefont {Williams},\ and\ \citenamefont
  {Ishigami}}]{ImpuritiesIshigami}%
  \BibitemOpen
  \bibfield  {author} {\bibinfo {author} {\bibfnamefont {J.-H.}\ \bibnamefont
  {Chen}}, \bibinfo {author} {\bibfnamefont {C.}~\bibnamefont {Jang}}, \bibinfo
  {author} {\bibfnamefont {S.}~\bibnamefont {Adam}}, \bibinfo {author}
  {\bibfnamefont {M.~S.}\ \bibnamefont {Fuhrer}}, \bibinfo {author}
  {\bibfnamefont {E.~D.}\ \bibnamefont {Williams}},\ and\ \bibinfo {author}
  {\bibfnamefont {M.}~\bibnamefont {Ishigami}},\ }\bibfield  {title} {\bibinfo
  {title} {Charged-impurity scattering in graphene},\ }\href
  {https://doi.org/10.1038/nphys935} {\bibfield  {journal} {\bibinfo  {journal}
  {Nature Physics}\ }\textbf {\bibinfo {volume} {4}},\ \bibinfo {pages} {377}
  (\bibinfo {year} {2008})}\BibitemShut {NoStop}%
\bibitem [{\citenamefont {Neumann}\ \emph {et~al.}(2015)\citenamefont
  {Neumann}, \citenamefont {Reichardt}, \citenamefont {Venezuela},
  \citenamefont {Dr{\"o}geler}, \citenamefont {Banszerus}, \citenamefont
  {Schmitz}, \citenamefont {Watanabe}, \citenamefont {Taniguchi}, \citenamefont
  {Mauri}, \citenamefont {Beschoten}, \citenamefont {Rotkin},\ and\
  \citenamefont {Stampfer}}]{StrainDisorderMapping}%
  \BibitemOpen
  \bibfield  {author} {\bibinfo {author} {\bibfnamefont {C.}~\bibnamefont
  {Neumann}}, \bibinfo {author} {\bibfnamefont {S.}~\bibnamefont {Reichardt}},
  \bibinfo {author} {\bibfnamefont {P.}~\bibnamefont {Venezuela}}, \bibinfo
  {author} {\bibfnamefont {M.}~\bibnamefont {Dr{\"o}geler}}, \bibinfo {author}
  {\bibfnamefont {L.}~\bibnamefont {Banszerus}}, \bibinfo {author}
  {\bibfnamefont {M.}~\bibnamefont {Schmitz}}, \bibinfo {author} {\bibfnamefont
  {K.}~\bibnamefont {Watanabe}}, \bibinfo {author} {\bibfnamefont
  {T.}~\bibnamefont {Taniguchi}}, \bibinfo {author} {\bibfnamefont
  {F.}~\bibnamefont {Mauri}}, \bibinfo {author} {\bibfnamefont
  {B.}~\bibnamefont {Beschoten}}, \bibinfo {author} {\bibfnamefont {S.~V.}\
  \bibnamefont {Rotkin}},\ and\ \bibinfo {author} {\bibfnamefont
  {C.}~\bibnamefont {Stampfer}},\ }\bibfield  {title} {\bibinfo {title} {Raman
  spectroscopy as probe of nanometre-scale strain variations in graphene},\
  }\href {https://doi.org/10.1038/ncomms9429} {\bibfield  {journal} {\bibinfo
  {journal} {Nature Communications}\ }\textbf {\bibinfo {volume} {6}},\
  \bibinfo {pages} {8429} (\bibinfo {year} {2015})}\BibitemShut {NoStop}%
\bibitem [{\citenamefont {Castro~Neto}\ \emph {et~al.}(2009)\citenamefont
  {Castro~Neto}, \citenamefont {Guinea}, \citenamefont {Peres}, \citenamefont
  {Novoselov},\ and\ \citenamefont {Geim}}]{RMPgraphene}%
  \BibitemOpen
  \bibfield  {author} {\bibinfo {author} {\bibfnamefont {A.~H.}\ \bibnamefont
  {Castro~Neto}}, \bibinfo {author} {\bibfnamefont {F.}~\bibnamefont {Guinea}},
  \bibinfo {author} {\bibfnamefont {N.~M.~R.}\ \bibnamefont {Peres}}, \bibinfo
  {author} {\bibfnamefont {K.~S.}\ \bibnamefont {Novoselov}},\ and\ \bibinfo
  {author} {\bibfnamefont {A.~K.}\ \bibnamefont {Geim}},\ }\bibfield  {title}
  {\bibinfo {title} {The electronic properties of graphene},\ }\href
  {https://doi.org/10.1103/RevModPhys.81.109} {\bibfield  {journal} {\bibinfo
  {journal} {Rev. Mod. Phys.}\ }\textbf {\bibinfo {volume} {81}},\ \bibinfo
  {pages} {109} (\bibinfo {year} {2009})}\BibitemShut {NoStop}%
\bibitem [{\citenamefont {Couto}\ \emph {et~al.}(2014)\citenamefont {Couto},
  \citenamefont {Costanzo}, \citenamefont {Engels}, \citenamefont {Ki},
  \citenamefont {Watanabe}, \citenamefont {Taniguchi}, \citenamefont
  {Stampfer}, \citenamefont {Guinea},\ and\ \citenamefont
  {Morpurgo}}]{grapheneStrainDisorderPRX}%
  \BibitemOpen
  \bibfield  {author} {\bibinfo {author} {\bibfnamefont {N.~J.~G.}\
  \bibnamefont {Couto}}, \bibinfo {author} {\bibfnamefont {D.}~\bibnamefont
  {Costanzo}}, \bibinfo {author} {\bibfnamefont {S.}~\bibnamefont {Engels}},
  \bibinfo {author} {\bibfnamefont {D.-K.}\ \bibnamefont {Ki}}, \bibinfo
  {author} {\bibfnamefont {K.}~\bibnamefont {Watanabe}}, \bibinfo {author}
  {\bibfnamefont {T.}~\bibnamefont {Taniguchi}}, \bibinfo {author}
  {\bibfnamefont {C.}~\bibnamefont {Stampfer}}, \bibinfo {author}
  {\bibfnamefont {F.}~\bibnamefont {Guinea}},\ and\ \bibinfo {author}
  {\bibfnamefont {A.~F.}\ \bibnamefont {Morpurgo}},\ }\bibfield  {title}
  {\bibinfo {title} {Random strain fluctuations as dominant disorder source for
  high-quality on-substrate graphene devices},\ }\href
  {https://doi.org/10.1103/PhysRevX.4.041019} {\bibfield  {journal} {\bibinfo
  {journal} {Phys. Rev. X}\ }\textbf {\bibinfo {volume} {4}},\ \bibinfo {pages}
  {041019} (\bibinfo {year} {2014})}\BibitemShut {NoStop}%
\bibitem [{\citenamefont {Shavit}\ \emph {et~al.}(2023)\citenamefont {Shavit},
  \citenamefont {Kol\'a\ifmmode~\check{r}\else \v{r}\fi{}}, \citenamefont
  {Mora}, \citenamefont {von Oppen},\ and\ \citenamefont
  {Oreg}}]{straindisordertbg}%
  \BibitemOpen
  \bibfield  {author} {\bibinfo {author} {\bibfnamefont {G.}~\bibnamefont
  {Shavit}}, \bibinfo {author} {\bibfnamefont {K.~c.~v.}\ \bibnamefont
  {Kol\'a\ifmmode~\check{r}\else \v{r}\fi{}}}, \bibinfo {author} {\bibfnamefont
  {C.}~\bibnamefont {Mora}}, \bibinfo {author} {\bibfnamefont {F.}~\bibnamefont
  {von Oppen}},\ and\ \bibinfo {author} {\bibfnamefont {Y.}~\bibnamefont
  {Oreg}},\ }\bibfield  {title} {\bibinfo {title} {Strain disorder and gapless
  intervalley coherent phase in twisted bilayer graphene},\ }\href
  {https://doi.org/10.1103/PhysRevB.107.L081403} {\bibfield  {journal}
  {\bibinfo  {journal} {Phys. Rev. B}\ }\textbf {\bibinfo {volume} {107}},\
  \bibinfo {pages} {L081403} (\bibinfo {year} {2023})}\BibitemShut {NoStop}%
\bibitem [{\citenamefont {Bolotin}\ \emph {et~al.}(2008)\citenamefont
  {Bolotin}, \citenamefont {Sikes}, \citenamefont {Jiang}, \citenamefont
  {Klima}, \citenamefont {Fudenberg}, \citenamefont {Hone}, \citenamefont
  {Kim},\ and\ \citenamefont {Stormer}}]{chargeinhomegeneitypeakwidth}%
  \BibitemOpen
  \bibfield  {author} {\bibinfo {author} {\bibfnamefont {K.}~\bibnamefont
  {Bolotin}}, \bibinfo {author} {\bibfnamefont {K.}~\bibnamefont {Sikes}},
  \bibinfo {author} {\bibfnamefont {Z.}~\bibnamefont {Jiang}}, \bibinfo
  {author} {\bibfnamefont {M.}~\bibnamefont {Klima}}, \bibinfo {author}
  {\bibfnamefont {G.}~\bibnamefont {Fudenberg}}, \bibinfo {author}
  {\bibfnamefont {J.}~\bibnamefont {Hone}}, \bibinfo {author} {\bibfnamefont
  {P.}~\bibnamefont {Kim}},\ and\ \bibinfo {author} {\bibfnamefont
  {H.}~\bibnamefont {Stormer}},\ }\bibfield  {title} {\bibinfo {title}
  {Ultrahigh electron mobility in suspended graphene},\ }\href
  {https://doi.org/https://doi.org/10.1016/j.ssc.2008.02.024} {\bibfield
  {journal} {\bibinfo  {journal} {Solid State Communications}\ }\textbf
  {\bibinfo {volume} {146}},\ \bibinfo {pages} {351} (\bibinfo {year}
  {2008})}\BibitemShut {NoStop}%
\bibitem [{\citenamefont {Dean}\ \emph {et~al.}(2010)\citenamefont {Dean},
  \citenamefont {Young}, \citenamefont {Meric}, \citenamefont {Lee},
  \citenamefont {Wang}, \citenamefont {Sorgenfrei}, \citenamefont {Watanabe},
  \citenamefont {Taniguchi}, \citenamefont {Kim}, \citenamefont {Shepard},\
  and\ \citenamefont {Hone}}]{Dean2010hBN}%
  \BibitemOpen
  \bibfield  {author} {\bibinfo {author} {\bibfnamefont {C.~R.}\ \bibnamefont
  {Dean}}, \bibinfo {author} {\bibfnamefont {A.~F.}\ \bibnamefont {Young}},
  \bibinfo {author} {\bibfnamefont {I.}~\bibnamefont {Meric}}, \bibinfo
  {author} {\bibfnamefont {C.}~\bibnamefont {Lee}}, \bibinfo {author}
  {\bibfnamefont {L.}~\bibnamefont {Wang}}, \bibinfo {author} {\bibfnamefont
  {S.}~\bibnamefont {Sorgenfrei}}, \bibinfo {author} {\bibfnamefont
  {K.}~\bibnamefont {Watanabe}}, \bibinfo {author} {\bibfnamefont
  {T.}~\bibnamefont {Taniguchi}}, \bibinfo {author} {\bibfnamefont
  {P.}~\bibnamefont {Kim}}, \bibinfo {author} {\bibfnamefont {K.~L.}\
  \bibnamefont {Shepard}},\ and\ \bibinfo {author} {\bibfnamefont
  {J.}~\bibnamefont {Hone}},\ }\bibfield  {title} {\bibinfo {title} {Boron
  nitride substrates for high-quality graphene electronics},\ }\href
  {https://doi.org/10.1038/nnano.2010.172} {\bibfield  {journal} {\bibinfo
  {journal} {Nature Nanotechnology}\ }\textbf {\bibinfo {volume} {5}},\
  \bibinfo {pages} {722} (\bibinfo {year} {2010})}\BibitemShut {NoStop}%
\bibitem [{\citenamefont {Zomer}\ \emph {et~al.}(2011)\citenamefont {Zomer},
  \citenamefont {Dash}, \citenamefont {Tombros},\ and\ \citenamefont {van
  Wees}}]{doi:10.1063/1.3665405hBN}%
  \BibitemOpen
  \bibfield  {author} {\bibinfo {author} {\bibfnamefont {P.~J.}\ \bibnamefont
  {Zomer}}, \bibinfo {author} {\bibfnamefont {S.~P.}\ \bibnamefont {Dash}},
  \bibinfo {author} {\bibfnamefont {N.}~\bibnamefont {Tombros}},\ and\ \bibinfo
  {author} {\bibfnamefont {B.~J.}\ \bibnamefont {van Wees}},\ }\bibfield
  {title} {\bibinfo {title} {A transfer technique for high mobility graphene
  devices on commercially available hexagonal boron nitride},\ }\href
  {https://doi.org/10.1063/1.3665405} {\bibfield  {journal} {\bibinfo
  {journal} {Applied Physics Letters}\ }\textbf {\bibinfo {volume} {99}},\
  \bibinfo {pages} {232104} (\bibinfo {year} {2011})},\ \Eprint
  {https://arxiv.org/abs/https://doi.org/10.1063/1.3665405}
  {https://doi.org/10.1063/1.3665405} \BibitemShut {NoStop}%
\bibitem [{\citenamefont {Dean}\ \emph {et~al.}(2011)\citenamefont {Dean},
  \citenamefont {Young}, \citenamefont {Cadden-Zimansky}, \citenamefont {Wang},
  \citenamefont {Ren}, \citenamefont {Watanabe}, \citenamefont {Taniguchi},
  \citenamefont {Kim}, \citenamefont {Hone},\ and\ \citenamefont
  {Shepard}}]{Dean2011hBN}%
  \BibitemOpen
  \bibfield  {author} {\bibinfo {author} {\bibfnamefont {C.~R.}\ \bibnamefont
  {Dean}}, \bibinfo {author} {\bibfnamefont {A.~F.}\ \bibnamefont {Young}},
  \bibinfo {author} {\bibfnamefont {P.}~\bibnamefont {Cadden-Zimansky}},
  \bibinfo {author} {\bibfnamefont {L.}~\bibnamefont {Wang}}, \bibinfo {author}
  {\bibfnamefont {H.}~\bibnamefont {Ren}}, \bibinfo {author} {\bibfnamefont
  {K.}~\bibnamefont {Watanabe}}, \bibinfo {author} {\bibfnamefont
  {T.}~\bibnamefont {Taniguchi}}, \bibinfo {author} {\bibfnamefont
  {P.}~\bibnamefont {Kim}}, \bibinfo {author} {\bibfnamefont {J.}~\bibnamefont
  {Hone}},\ and\ \bibinfo {author} {\bibfnamefont {K.~L.}\ \bibnamefont
  {Shepard}},\ }\bibfield  {title} {\bibinfo {title} {Multicomponent fractional
  quantum hall effect in graphene},\ }\href
  {https://doi.org/10.1038/nphys2007} {\bibfield  {journal} {\bibinfo
  {journal} {Nature Physics}\ }\textbf {\bibinfo {volume} {7}},\ \bibinfo
  {pages} {693} (\bibinfo {year} {2011})}\BibitemShut {NoStop}%
\bibitem [{\citenamefont {Dong}\ \emph {et~al.}(2023)\citenamefont {Dong},
  \citenamefont {Chubukov},\ and\ \citenamefont
  {Levitov}}]{ChubukovLevitovAlternative}%
  \BibitemOpen
  \bibfield  {author} {\bibinfo {author} {\bibfnamefont {Z.}~\bibnamefont
  {Dong}}, \bibinfo {author} {\bibfnamefont {A.~V.}\ \bibnamefont {Chubukov}},\
  and\ \bibinfo {author} {\bibfnamefont {L.}~\bibnamefont {Levitov}},\
  }\bibfield  {title} {\bibinfo {title} {Transformer spin-triplet
  superconductivity at the onset of isospin order in bilayer graphene},\ }\href
  {https://doi.org/10.1103/PhysRevB.107.174512} {\bibfield  {journal} {\bibinfo
   {journal} {Phys. Rev. B}\ }\textbf {\bibinfo {volume} {107}},\ \bibinfo
  {pages} {174512} (\bibinfo {year} {2023})}\BibitemShut {NoStop}%
\bibitem [{\citenamefont {Wagner}\ \emph {et~al.}(2023)\citenamefont {Wagner},
  \citenamefont {Kwan}, \citenamefont {Bultinck}, \citenamefont {Simon},\ and\
  \citenamefont {Parameswaran}}]{KWANBLG}%
  \BibitemOpen
  \bibfield  {author} {\bibinfo {author} {\bibfnamefont {G.}~\bibnamefont
  {Wagner}}, \bibinfo {author} {\bibfnamefont {Y.~H.}\ \bibnamefont {Kwan}},
  \bibinfo {author} {\bibfnamefont {N.}~\bibnamefont {Bultinck}}, \bibinfo
  {author} {\bibfnamefont {S.~H.}\ \bibnamefont {Simon}},\ and\ \bibinfo
  {author} {\bibfnamefont {S.~A.}\ \bibnamefont {Parameswaran}},\ }\href
  {https://doi.org/10.48550/ARXIV.2302.00682} {\bibinfo {title}
  {Superconductivity from repulsive interactions in bernal-stacked bilayer
  graphene}} (\bibinfo {year} {2023})\BibitemShut {NoStop}%
\bibitem [{\citenamefont {Jimeno-Pozo}\ \emph {et~al.}(2023)\citenamefont
  {Jimeno-Pozo}, \citenamefont {Sainz-Cruz}, \citenamefont {Cea}, \citenamefont
  {Pantale\'on},\ and\ \citenamefont {Guinea}}]{AlejandroPacoBLG}%
  \BibitemOpen
  \bibfield  {author} {\bibinfo {author} {\bibfnamefont {A.}~\bibnamefont
  {Jimeno-Pozo}}, \bibinfo {author} {\bibfnamefont {H.}~\bibnamefont
  {Sainz-Cruz}}, \bibinfo {author} {\bibfnamefont {T.}~\bibnamefont {Cea}},
  \bibinfo {author} {\bibfnamefont {P.~A.}\ \bibnamefont {Pantale\'on}},\ and\
  \bibinfo {author} {\bibfnamefont {F.}~\bibnamefont {Guinea}},\ }\bibfield
  {title} {\bibinfo {title} {Superconductivity from electronic interactions and
  spin-orbit enhancement in bilayer and trilayer graphene},\ }\href
  {https://doi.org/10.1103/PhysRevB.107.L161106} {\bibfield  {journal}
  {\bibinfo  {journal} {Phys. Rev. B}\ }\textbf {\bibinfo {volume} {107}},\
  \bibinfo {pages} {L161106} (\bibinfo {year} {2023})}\BibitemShut {NoStop}%
\bibitem [{\citenamefont {Chou}\ \emph
  {et~al.}(2022{\natexlab{b}})\citenamefont {Chou}, \citenamefont {Wu},\ and\
  \citenamefont {Das~Sarma}}]{dassarmaBLGsUBSTARTEiNDUCEDEFFECTS}%
  \BibitemOpen
  \bibfield  {author} {\bibinfo {author} {\bibfnamefont {Y.-Z.}\ \bibnamefont
  {Chou}}, \bibinfo {author} {\bibfnamefont {F.}~\bibnamefont {Wu}},\ and\
  \bibinfo {author} {\bibfnamefont {S.}~\bibnamefont {Das~Sarma}},\ }\bibfield
  {title} {\bibinfo {title} {Enhanced superconductivity through virtual
  tunneling in bernal bilayer graphene coupled to ${\mathrm{wse}}_{2}$},\
  }\href {https://doi.org/10.1103/PhysRevB.106.L180502} {\bibfield  {journal}
  {\bibinfo  {journal} {Phys. Rev. B}\ }\textbf {\bibinfo {volume} {106}},\
  \bibinfo {pages} {L180502} (\bibinfo {year}
  {2022}{\natexlab{b}})}\BibitemShut {NoStop}%
\bibitem [{\citenamefont {Chou}\ \emph
  {et~al.}(2022{\natexlab{c}})\citenamefont {Chou}, \citenamefont {Wu},
  \citenamefont {Sau},\ and\ \citenamefont
  {Das~Sarma}}]{DasSarmaPhonosGraphene}%
  \BibitemOpen
  \bibfield  {author} {\bibinfo {author} {\bibfnamefont {Y.-Z.}\ \bibnamefont
  {Chou}}, \bibinfo {author} {\bibfnamefont {F.}~\bibnamefont {Wu}}, \bibinfo
  {author} {\bibfnamefont {J.~D.}\ \bibnamefont {Sau}},\ and\ \bibinfo {author}
  {\bibfnamefont {S.}~\bibnamefont {Das~Sarma}},\ }\bibfield  {title} {\bibinfo
  {title} {Acoustic-phonon-mediated superconductivity in moir\'eless graphene
  multilayers},\ }\href {https://doi.org/10.1103/PhysRevB.106.024507}
  {\bibfield  {journal} {\bibinfo  {journal} {Phys. Rev. B}\ }\textbf {\bibinfo
  {volume} {106}},\ \bibinfo {pages} {024507} (\bibinfo {year}
  {2022}{\natexlab{c}})}\BibitemShut {NoStop}%
\bibitem [{\citenamefont {Ghazaryan}\ \emph {et~al.}(2023)\citenamefont
  {Ghazaryan}, \citenamefont {Holder}, \citenamefont {Berg},\ and\
  \citenamefont {Serbyn}}]{BErgHolderPrbkohnluttingergraphene}%
  \BibitemOpen
  \bibfield  {author} {\bibinfo {author} {\bibfnamefont {A.}~\bibnamefont
  {Ghazaryan}}, \bibinfo {author} {\bibfnamefont {T.}~\bibnamefont {Holder}},
  \bibinfo {author} {\bibfnamefont {E.}~\bibnamefont {Berg}},\ and\ \bibinfo
  {author} {\bibfnamefont {M.}~\bibnamefont {Serbyn}},\ }\bibfield  {title}
  {\bibinfo {title} {Multilayer graphenes as a platform for interaction-driven
  physics and topological superconductivity},\ }\href
  {https://doi.org/10.1103/PhysRevB.107.104502} {\bibfield  {journal} {\bibinfo
   {journal} {Phys. Rev. B}\ }\textbf {\bibinfo {volume} {107}},\ \bibinfo
  {pages} {104502} (\bibinfo {year} {2023})}\BibitemShut {NoStop}%
\bibitem [{\citenamefont {Ghazaryan}\ \emph {et~al.}(2021)\citenamefont
  {Ghazaryan}, \citenamefont {Holder}, \citenamefont {Serbyn},\ and\
  \citenamefont {Berg}}]{annularFermiSC}%
  \BibitemOpen
  \bibfield  {author} {\bibinfo {author} {\bibfnamefont {A.}~\bibnamefont
  {Ghazaryan}}, \bibinfo {author} {\bibfnamefont {T.}~\bibnamefont {Holder}},
  \bibinfo {author} {\bibfnamefont {M.}~\bibnamefont {Serbyn}},\ and\ \bibinfo
  {author} {\bibfnamefont {E.}~\bibnamefont {Berg}},\ }\bibfield  {title}
  {\bibinfo {title} {Unconventional superconductivity in systems with annular
  fermi surfaces: Application to rhombohedral trilayer graphene},\ }\href
  {https://doi.org/10.1103/PhysRevLett.127.247001} {\bibfield  {journal}
  {\bibinfo  {journal} {Phys. Rev. Lett.}\ }\textbf {\bibinfo {volume} {127}},\
  \bibinfo {pages} {247001} (\bibinfo {year} {2021})}\BibitemShut {NoStop}%
\bibitem [{\citenamefont {You}\ and\ \citenamefont
  {Vishwanath}(2022)}]{AshvinRTGIVC}%
  \BibitemOpen
  \bibfield  {author} {\bibinfo {author} {\bibfnamefont {Y.-Z.}\ \bibnamefont
  {You}}\ and\ \bibinfo {author} {\bibfnamefont {A.}~\bibnamefont
  {Vishwanath}},\ }\bibfield  {title} {\bibinfo {title} {Kohn-luttinger
  superconductivity and intervalley coherence in rhombohedral trilayer
  graphene},\ }\href {https://doi.org/10.1103/PhysRevB.105.134524} {\bibfield
  {journal} {\bibinfo  {journal} {Phys. Rev. B}\ }\textbf {\bibinfo {volume}
  {105}},\ \bibinfo {pages} {134524} (\bibinfo {year} {2022})}\BibitemShut
  {NoStop}%
\bibitem [{\citenamefont {Chatterjee}\ \emph {et~al.}(2022)\citenamefont
  {Chatterjee}, \citenamefont {Wang}, \citenamefont {Berg},\ and\ \citenamefont
  {Zaletel}}]{ChatterjeeIVCRTG}%
  \BibitemOpen
  \bibfield  {author} {\bibinfo {author} {\bibfnamefont {S.}~\bibnamefont
  {Chatterjee}}, \bibinfo {author} {\bibfnamefont {T.}~\bibnamefont {Wang}},
  \bibinfo {author} {\bibfnamefont {E.}~\bibnamefont {Berg}},\ and\ \bibinfo
  {author} {\bibfnamefont {M.~P.}\ \bibnamefont {Zaletel}},\ }\bibfield
  {title} {\bibinfo {title} {Inter-valley coherent order and isospin
  fluctuation mediated superconductivity in rhombohedral trilayer graphene},\
  }\href {https://doi.org/10.1038/s41467-022-33561-w} {\bibfield  {journal}
  {\bibinfo  {journal} {Nature Communications}\ }\textbf {\bibinfo {volume}
  {13}},\ \bibinfo {pages} {6013} (\bibinfo {year} {2022})}\BibitemShut
  {NoStop}%
\bibitem [{\citenamefont {Jamei}\ \emph {et~al.}(2005)\citenamefont {Jamei},
  \citenamefont {Kivelson},\ and\ \citenamefont
  {Spivak}}]{KivelsonSpivakFirstOrder}%
  \BibitemOpen
  \bibfield  {author} {\bibinfo {author} {\bibfnamefont {R.}~\bibnamefont
  {Jamei}}, \bibinfo {author} {\bibfnamefont {S.}~\bibnamefont {Kivelson}},\
  and\ \bibinfo {author} {\bibfnamefont {B.}~\bibnamefont {Spivak}},\
  }\bibfield  {title} {\bibinfo {title} {Universal aspects of
  coulomb-frustrated phase separation},\ }\href
  {https://doi.org/10.1103/PhysRevLett.94.056805} {\bibfield  {journal}
  {\bibinfo  {journal} {Phys. Rev. Lett.}\ }\textbf {\bibinfo {volume} {94}},\
  \bibinfo {pages} {056805} (\bibinfo {year} {2005})}\BibitemShut {NoStop}%
\end{thebibliography}%
\end{document}